\documentclass[12pt]{spieman}
\usepackage{amsmath,amsfonts,amssymb}
\usepackage{graphicx}
\usepackage{setspace}
\usepackage{tocloft}
\usepackage{lineno}
\usepackage{algorithmic}
\usepackage{algorithm, setspace}
\usepackage{subcaption}
\usepackage{multirow}
\usepackage{bm}
\usepackage{diagbox}
\usepackage[labelfont=bf, skip=8pt]{caption}

\usepackage{fancyhdr}
\usepackage{booktabs}

\usepackage[colorlinks=true, allcolors=blue]{hyperref}

\newcommand{\R}{\mathbb{R}}
\newcommand{\E}{\mathbb{E}}

\newcommand{\x}{\bm{x}}
\newcommand{\br}{\bm{r}}
\newcommand{\bv}{\bm{v}}
\newcommand{\Var}{\text{Var}}

\def\PAnumber{2025-5580}
\fancypagestyle{empty}{%
  \fancyhf{} 
  \fancyfoot[C]{\fontsize{11}{11}\selectfont Approved for public release; distribution is unlimited. Public Affairs release approval \# \PAnumber .}
  
}

\title{Boiling flow estimation for aero-optic phase screen generation}

\author[a,*]{Jeffrey W. Utley}
\author[a]{Gregery T. Buzzard}
\author[b]{Charles A. Bouman}
\author[c]{Matthew R. Kemnetz}
\affil[a]{Purdue University, Department of Mathematics, West Lafayette, Indiana 47907, USA}
\affil[b]{Purdue University, Departments of Electrical and Computer Engineering and Biomedical Engineering, West Lafayette, Indiana 47907, USA}
\affil[c]{Air Force Institute of Technology, Department of Engineering Physics, Wright-Patterson AFB, OH 45433, USA}

\cftpagenumbersoff{figure}
\cftpagenumbersoff{table} 
\begin{document} 
\maketitle

\begin{abstract}
Aero-optic effects due to turbulence can reduce the effectiveness of transmitting light waves to a distant target. Methods to compensate for turbulence typically rely on realistic turbulence data, which can be generated by i) experiment, ii) high-fidelity CFD, iii) low-fidelity CFD, and iv) autoregressive methods. However, each of these methods has significant drawbacks, including monetary and/or computational expense, limited quantity, inaccurate statistics, and overall complexity. In contrast, the boiling flow algorithm is a simple, computationally efficient model that can generate atmospheric phase screen data with only a handful of parameters.  However, boiling flow has not been widely used in aero-optic applications, at least in part because some of these parameters, such as $r_0$, are not clearly defined for aero-optic data. In this paper, we demonstrate a method to use the boiling flow algorithm to generate arbitrary length synthetic data to match the statistics of measured aero-optic data. Importantly, we modify the standard boiling flow method to generate anisotropic phase screens. While this model does not fully capture all statistics, it can be used to generate data that matches the temporal power spectrum or the anisotropic 2D structure function, with the ability to trade fidelity to one for fidelity to the other.  
\end{abstract}

\keywords{Aero-optics, turbulent boundary layer, phase screens, wavefront aberrations, frozen flow, boiling flow}

{\noindent \footnotesize\textbf{*}Jeffrey W. Utley,  \linkable{utleyj@purdue.edu} }

\begin{spacing}{2}

\section{Introduction}
\label{s: Introduction}

Atmospheric and aerodynamic turbulence distort light wave propagation, thereby reducing the effectiveness of transmitting light waves to a distant target. In particular, both phenomena cause refractive index variations \cite{Tatarski, WangPhysicsComputation, Fitzgerald} which lead to random phase aberrations \cite{Visbal, Chernov, Kalensky}. Here, we focus on the phase aberrations induced by aerodynamic turbulence, which are called aero-optic phase aberrations. These spatially and temporally varying phase aberrations can be measured by optical sensors \cite{Kemnetz, Geary, Holmes}, resulting in a time series of images called aero-optic phase screens. Modern methods to mitigate these aberrations include algorithms such as dynamic mode decomposition \cite{Shaffer, Kutz, ShafferPredictive, Sahba}, machine learning \cite{ShafferNeuralNetwork, BurnsALatency, BurnsARobust}, and autoregressive modeling \cite{BurnsEstimation}. However, each of these methods requires long time series of aero-optic phase screen data to yield adequate training without overfitting in these algorithms.

Measuring long sequences of aero-optic phase screens through experiment is expensive and time intensive \cite{JumperAAOL}, so a variety of simulation methods have been proposed as a less expensive alternative. Computational Fluid Dynamics (CFD) can simulate an aerodynamic flow field \cite{JumperPhysicsMeasurement, WangPhysicsComputation, Visbal}, from which one can derive the resulting phase aberrations \cite{WangAero-Optics, WangComputation, Porter}. High-fidelity CFD can accurately model the aerodynamic flow field around well-understood geometries \cite{WangComputation, WangAero-Optics, Visbal} but is very computationally expensive \cite{GordeyevFluidDynamics, WangPhysicsComputation, GordeyevFluidDynamics2}, while low-fidelity CFD simplifies the complex calculations involved, but gives less accurate results \cite{GordeyevFluidDynamics, WangPhysicsComputation}. In contrast, the autoregressive methods introduced by Vogel et al. \cite{Vogel}, Utley et al. \cite{Utley}, and Faghihi et al. \cite{Faghihi} are less computationally expensive and match the statistics of measured aero-optic data. However, these methods use complex statistical models with many unknown parameters. 

\begin{figure}[t]
    \centering
    \includegraphics[width=0.7 \textwidth]{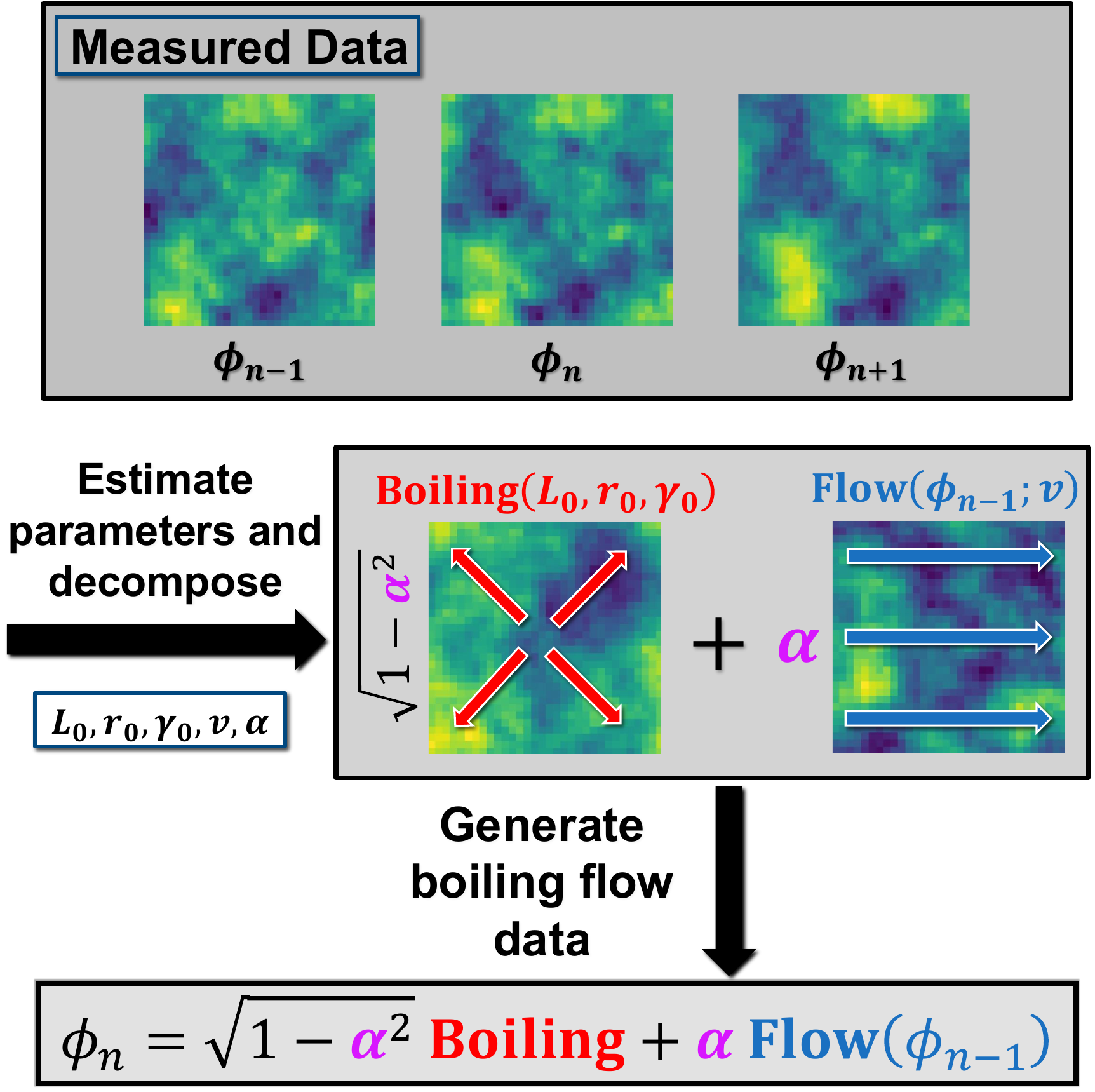}
    \caption{Outline of the boiling flow algorithm as used in this paper. Starting with a time series of measured phase screens $\phi_n$, our method estimates the boiling flow parameters and decomposes the data into the weighted sum of a boiling component and a flow component. The method then uses these components and the parameters to generate synthetic phase screens using boiling flow. We introduce the parameter $\gamma_0$ into the boiling model to produce spatially anisotropic correlations.}
    \label{fig: Overview}
\end{figure}
 
Although not used previously for generating aero-optic phase screens, the simulation algorithm called boiling flow \cite{Srinath} is a computationally efficient algorithm for generating physically-relevant atmospheric phase screens using only five parameters. Introduced in Srinath et al. \cite{Srinath}, boiling flow generalizes a simulation method using the Kolmogorov theory of turbulence\cite{Schmidt, Srinath, Tatarski} and the Taylor frozen-flow hypothesis \cite{Taylor, Schmidt, Srinath, Vogel, PoyneerExperimental}.  Boiling flow has become a common and well-known method for generating atmospheric phase screens \cite{Gerard, Dayton, Lin, PoyneerLaboratory, Jia, Boddeda, Sheikh, Snyder, Sridhar, Lehtonen, eBraga, Lloyd, SheikhDynamic}. 
However, boiling flow depends on physical parameters that follow from the Kolmogorov theory of turbulence and the Taylor frozen-flow hypothesis.  Moreover, boiling flow produces data with spatially isotropic correlations. Since aero-optic effects do not follow the Kolmogorov theory of turbulence \cite{Vogel, Siegenthaler} and are spatially anisotropic, this approach has not been used for aero-optic phase screen generation.

In this paper, we introduce a method to use the boiling flow algorithm to generate arbitrary length, spatially anisotropic synthetic data to match the statistics of measured aero-optic data. Our key contributions are as follows:
\begin{itemize}
    \item Instead of interpreting boiling flow parameters as physical quantities, we introduce an algorithm that estimates these parameters to fit the temporal power spectrum of measured data.
    \item We introduce a new parameter $\gamma_0$ to the boiling model to produce spatially anisotropic synthetic data, which can be used to match the 2D structure function of measured data.
\end{itemize} 
Our results show that this method matches the temporal power spectrum (TPS) of measured aero-optic data within 12\% when using the standard, isotropic form of boiling and provides a good qualitative fit to the 2D structure function when using the new, anisotropic version. 

This paper builds on previous work from Ref.~\citenum{UtleyBoiling}. This paper uses the GitHub repository ``Boiling Flow," linked in Ref.~\citenum{Repo}. 

\section{Overview of Boiling Flow}
\label{s: Boiling Flow}

Figure~\ref{fig: Overview} shows an outline of boiling flow as used in this paper. Boiling flow depends on physical parameters that follow from the Kolmogorov theory of turbulence and the Taylor frozen-flow hypothesis. These parameters include the wind velocity components $(v_x,v_y)$, outer scale $L_0$, Fried coherence length $r_0$, and a flow-coefficient $\alpha$. Furthermore, here we introduce an anisotropy parameter $\gamma_0$ to adjust the spatial correlation length scale for the $y$-axis separately from the $x$-axis.  

The key idea of boiling flow, as described in greater detail in Ref.~\citenum{Srinath}, is to represent a sequence of phase screens in terms of two distinct components: boiling and flow.

Boiling is a random process with a specified power spectrum associated with Kolmogorov turbulence. 
In practice, a boiling phase screen is generated in Fourier space following Ref.~\citenum{Schmidt}.  
First define $\Delta_f = 1 / (N \Delta)$, where $N$ is the number of pixels along a side in a square phase screen and $\Delta$ is the spatial grid spacing in meters per pixel.  Then in Fourier space,
\begin{align} \label{eq: Boiling}
    \tilde{B}_n &= \Delta_f \sqrt{V_{\phi}\left(\bm{k} \Delta_f \right)}\;\odot \;\epsilon_n \hspace{1cm} \text{(Boiling at step $n$)}
\end{align}
Here, $\bm{k}$ is an array of 2-component entries, with each component taking integer values from $-N/2$ to $N/2-1$.  Also, $\epsilon_n$ is a 2D-array of complex-valued white noise with variance 1 in each component, and $V_\phi$ is the idealized Von K\'arman spatial power spectral density (PSD) defined in Eq.~\eqref{eq: Von Karman PSD}, with input in cycles per meter. $\tilde{B}_n$ depends on $L_0$, $\gamma_0$, and $r_0$, which determine the PSD through $V_\phi$.  Also, $\odot$ represents element-wise multiplication.

In contrast, flow is a purely convective term that translates a given phase screen according to the velocity vector $\bm{v} = (v_x, v_y)$.  
Taking $\tilde{\phi}_{n-1}$ to be a phase screen in Fourier space at time-step $n-1$, this gives a translated screen at time-step $n$
\begin{align} \label{eq: Flow}
    \tilde{F}_n = e^{-j (\bm{v} \cdot 2 \pi \bm{\bm{k}} / N)}\:\odot \:\tilde{\phi}_{n-1}. \hspace{1cm} \text{(Flow from $n-1$ to $n$)}
\end{align}
We use $F$ instead of $\phi$ to indicate that this is the flow component only; this will be combined with the boiling component in Eq. \eqref{eq: Boiling Flow}. Since the flow is represented in Fourier space, the translation is accomplished by multiplication. $\tilde{F}_n$ depends on the flow velocity $\bm{v}$.

Boiling flow uses an energy-preserving linear combination of these individual terms to model a convective motion that evolves at each step to include the effects of boiling.  
This yields the Fourier space representation of the phase screen at time step $n$ as
\begin{align}\label{eq: Boiling Flow}
    \tilde{\phi}_n &= 
    \sqrt{1-\alpha^2}\:\tilde{B}_n
    + 
    \alpha \:\tilde{F}_n
    . \hspace{1cm}  \text{(Boiling Flow)}
\end{align}
The scaling by $\alpha$ and $\sqrt{1 - \alpha^2}$ in Eq.~(\ref{eq: Boiling Flow}) ensures that the spatial PSD of each phase screen $\phi_n\in\R^{N\times N}$ is the idealized PSD $V_\phi$. 

Our generation algorithm starts with a random phase screen $\tilde{\phi}_0 = \tilde{B}_0$ and applies Eq.~(\ref{eq: Boiling Flow}) recursively, with the following modifications:
\begin{itemize}
    \item First, to reduce the effect of periodicity arising from a circular Fast Fourier Transform (FFT), we generate phase screens of size $4N\times 4N$ and then extract the first $N\times N$ indices from each over-sized screen.

    \item Second, we remove tilt, tip, and piston (TTP) from each (restricted) phase screen $\phi_n\in\R^{N\times N}$. This is a common practice for post-processing measured phase screen data \cite{Kemnetz, KemnetzAnalysis, SiegenthalerShear}, so this step improves the physical relevance of the synthetic phase screens.
\end{itemize}
This method provides a way to generate arbitrary-duration time series of physically-relevant phase screens.

\section{Boiling Flow Parameter Estimation}
\label{s: Parameter Estimation from Training Data}
Section~\ref{s: Boiling Flow} showed that the primary parameters of boiling flow are $L_0$, $\gamma_0$, and $r_0$ to describe boiling, $v_x$ and $v_y$ to determine the flow velocity, and $\alpha$ to determine the ratio of the boiling term to the flow term. 
Many applications of boiling flow select these parameters based on a desired physical environment \cite{Gerard, Dayton, Jia, Boddeda, Sheikh, Lehtonen, eBraga}, where $L_0$, $r_0$, and the magnitude of $(v_x,v_y)$ are derived from a combination of the Kolmogorov theory of turbulence \cite{Schmidt, Coulman, Fried} and the Taylor frozen-flow hypothesis \cite{Tyson, Sheikh}. 
However, because aero-optic effects do not follow the Kolmogorov theory of turbulence \cite{Vogel, Siegenthaler}, we need to develop and justify an approach to estimate these parameters from measured aero-optic phase screen data.

Existing algorithms designed for atmospheric turbulence could be used to estimate $(L_0, r_0)$ from measured phase aberration data. For example, Refs.~\citenum{Avila, ZiadFrom, Ziad} and Ref.~\citenum{Martin} estimate $L_0$ and $r_0$ (respectively) from the spatial covariance of the angle-of-arrival (AA) measurements. However, these models are designed for data with non-zero tip/tilt terms, whereas tip/tilt are generally removed from measured aero-optic phase screen data \cite{Kemnetz, KemnetzAnalysis, SiegenthalerShear}. In contrast, Ref.~\citenum{Schöck} presents two methods to estimate $(L_0, r_0)$ from tip/tilt-corrected phase screen data: (1) fitting the structure function and (2) fitting the Zernike modes of AA measurements. However, the former approach uses a grid-search algorithm for chosen values of $L_0$ and the latter method involves complex calculations (e.g., matrix inversion, singular value decomposition) to take Zernike modes with a non-circular aperture. Similarly, Ref.~\citenum{Andrade} estimates $(L_0, r_0)$ from the Zernike modes of phase screens, but uses an iterative algorithm with multiple least-squares calculations to solve an optimization problem which is not shown be convex. 

One existing method for estimating the velocity components $(v_x,v_y)$ is Poyneer et al.\cite{PoyneerExperimental}, which uses a heuristic grid search over velocity vectors. An extension of this method to estimate $\alpha$ using the width of the temporal power spectrum (TPS) peak(s) is found in Refs.~\citenum{Srinath, PoyneerLaboratory}. However, the basic method is designed for frozen flow with multiple wind layers rather than the single layer setting considered here.

In this section, we estimate the boiling flow parameters from a training data set containing a time series of $M\times N$ images. We note that Eq.~(\ref{eq: Boiling Flow}) is not linear in these parameters, so it is difficult to construct a convex loss function that depends on all parameters. Instead, we estimate each parameter with appropriate methods. 
\begin{itemize}
    \item We set $L_0$ to the width (in meters) of the aperture.
    \item For the isotropic boiling model, we use per-frequency scaling by $(r_0(\bm{f}))^{-5/3}$ to match the unit-scale Von K\'arman PSD\cite{Schmidt} to the spatial PSD of the measured data, then average over frequency to estimate the value of $r_0$ (in meters).
    \item For the anisotropic boiling model, we fix a value of $\gamma_0$ (unitless), then estimate $r_0 = r_0(\gamma_0)$ as above.  Then we choose $\gamma_0$ to minimize the mean-squared error between the 2D structure function of the resulting boiling and the 2D structure function of the data.   
    \item We find the flow velocity $\bm{v}$ (in pixels per time-step) that maximizes the spatial cross-correlation between TTP-removed frames at multiple time lags, while accounting for the uncertainty induced by boiling.  
    \item We find the flow-coefficient $\alpha$ (unitless) by minimizing the mean-squared error in Fourier space between the measured $\tilde{\phi}_n$ and the predicted flow $\alpha \tilde{F}_n$, obtained as in Eq.~\eqref{eq: Flow} by flowing one time step starting from $\tilde{\phi}_{n-1}$.  
\end{itemize}   
A key challenge in the estimation of $\bm{v}$ is the tension between using a longer time lag to better estimate the flow and the confounding effect of boiling, which decreases the correlation between screens separated by long time lags. 
We describe this challenge and details of each estimation procedure in further detail below.

\subsection{Estimating the Boiling Parameters}
\label{s: Estimation of Boiling Parameters}

Estimation of the boiling parameters $L_0$ and $r_0$ begins with understanding their role in the boiling model. The outer scale $L_0$ is defined as the upper bound of the size of turbulent eddies in the atmosphere \cite{Schmidt}.  However, the effect of turbulent eddies larger than the aperture is minimal in tip/tilt/piston (TTP)-removed phase screen data, which is common for aero-optic data \cite{Kemnetz, KemnetzAnalysis, SiegenthalerShear}.  Also, in the case of isotropic turbulence, the Fried coherence length $r_0$ is defined as the diameter over which the mean-squared phase error is at most one radian \cite{Schmidt, Fried}. However, aero-optic phase aberrations are typically anisotropic \cite{Vogel, Siegenthaler}, and $r_0$ alone does not capture this anisotropy. Thus, both parameters require adaptation to measured TTP-removed aero-optic data.   

\begin{figure}[t]
    \centering
    \begin{minipage}[t]{0.48\textwidth}
        \centering
        \vfill
        \includegraphics[width=\textwidth]{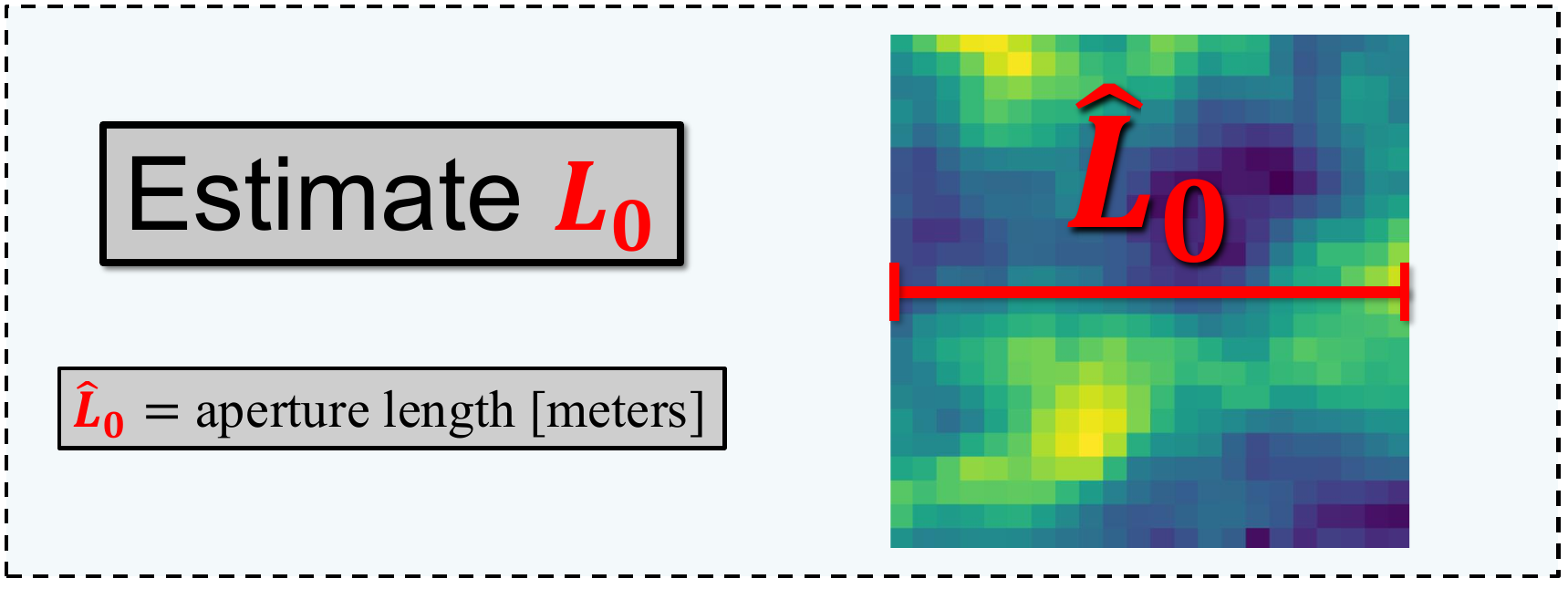}\\
        \includegraphics[width=\textwidth]{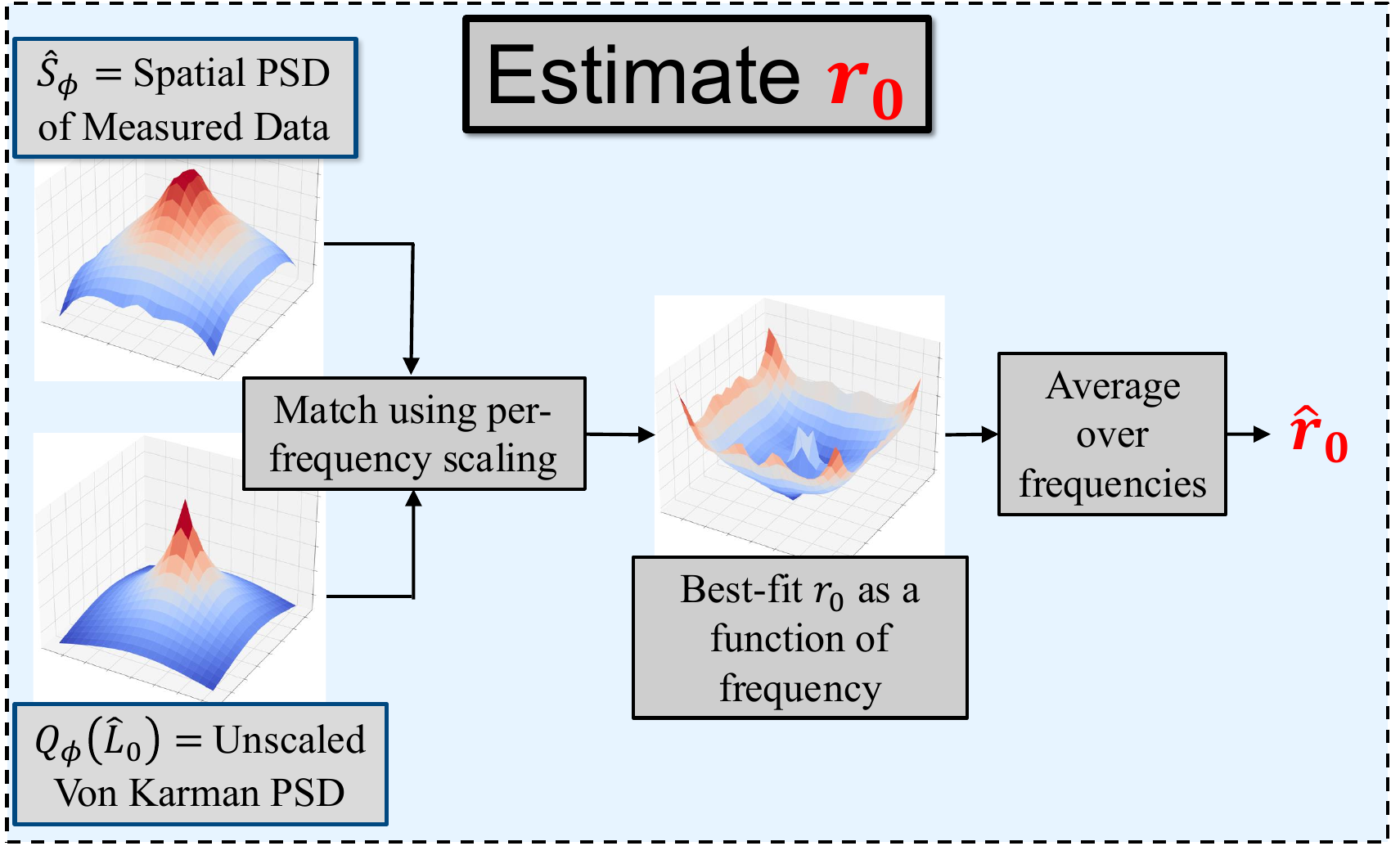}
    \end{minipage}
    \hfill
    \begin{minipage}[t]{0.48\textwidth}
        \centering
        \vfill
        \includegraphics[width=\textwidth]{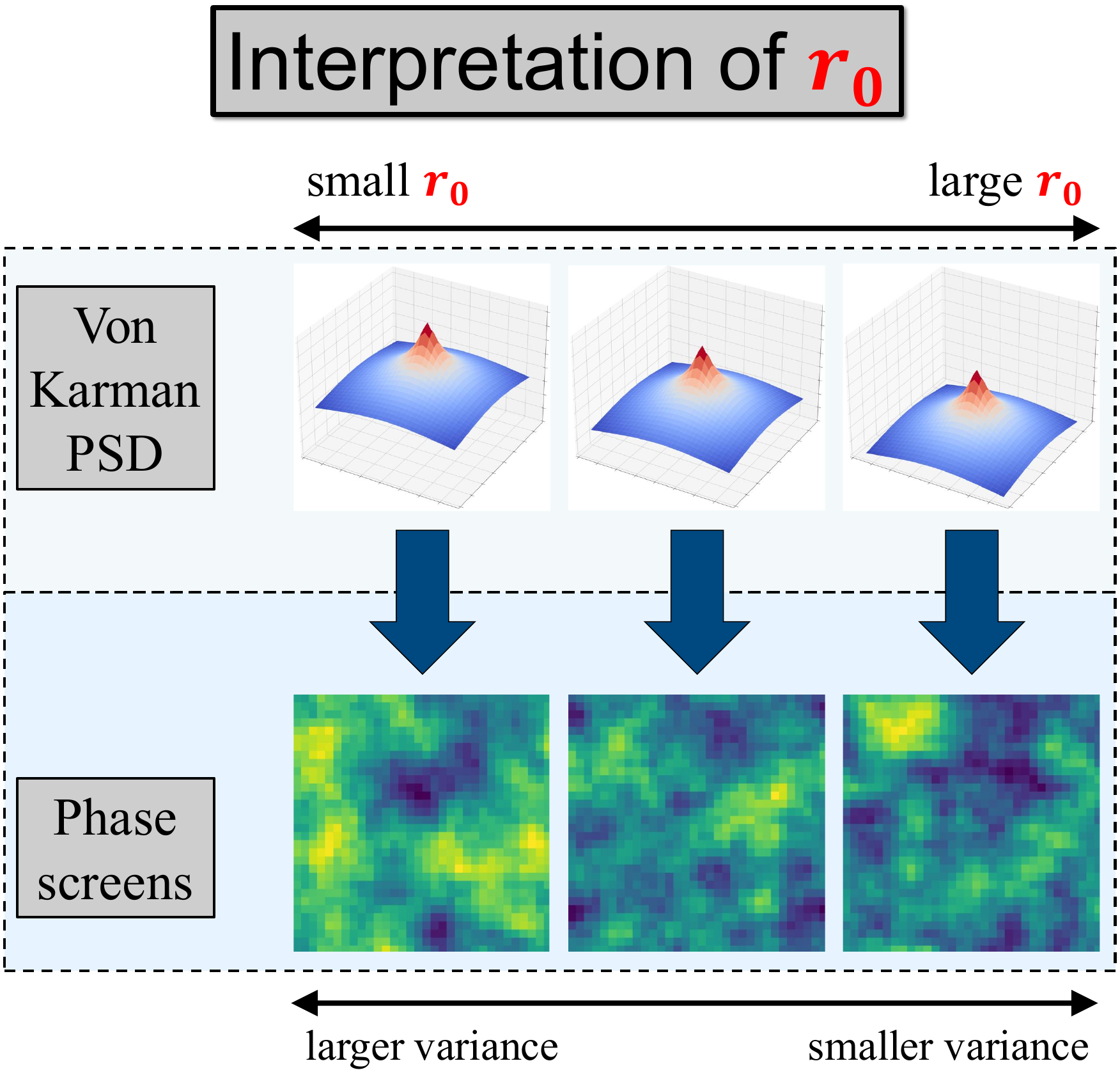}
    \end{minipage}

    \caption{{\bf Left:} Estimation of the boiling parameters $(L_0, r_0)$. We set $\hat{L}_0$ to the aperture length in meters. We then do per-frequency scaling by $(r_0(\bm{f}))^{-5/3}$ of the unit-scale Von K\'arman PSD to match the spatial PSD of the measured data. Then we average over frequencies to obtain $\hat{r}_0$. 
    {\bf Right:}  Instead of regarding $r_0$ as a length scale, we interpret it as a scaling parameter for the spatial PSD of phase screens. Smaller $r_0$ corresponds to larger variance of the phase screens.}
    \label{fig: Estimate Boiling Parameters}

\end{figure}

Figure~\ref{fig: Estimate Boiling Parameters}, left, illustrates our algorithm for estimating the boiling parameters $(L_0, r_0)$ to fit an isotropic model to the spatial PSD of the training data.
Figure~\ref{fig: Estimate Boiling Parameters}, right, describes our interpretation of the Fried coherence length $r_0$ in terms of the spatial PSD.

To estimate $L_0$, we note first that  
because turbulent eddies larger than the aperture mostly result in tip/tilt aberrations instead of higher-order phase aberrations \cite{WangPhysicsComputation}, removing tip/tilt from phase screen data significantly reduces the effects on large eddies. Further, Siegenthaler et al. \cite{SiegenthalerShear} show that tip/tilt removal restricts the (temporal) period of phase aberration data to be at most the aperture size; under the frozen-flow hypothesis, this period is proportional to the size of turbulent eddies\cite{Schmidt}. Thus, we set $\hat{L}_0$ to the length of the aperture of the measured data, which is the upper bound of the size of turbulent eddies that have a meaningful effect on the measured phase screen data.

To estimate $r_0$ in a way that generalizes to the anisotropic case, we consider the effect of $r_0$ on the Von K\'arman PSD, which is commonly used to generate spatially isotropic random phase screens \cite{Schmidt}.  As described more precisely below and illustrated in Fig.~\ref{fig: Estimate Boiling Parameters}, we first do a frequency dependent scaling of the Von K\'arman PSD to match the spatial PSD of the measured data, then average over frequency to estimate $r_0$. 

To extend this to the anisotropic case, we introduce a unitless parameter $\gamma_0$ to modulate the correlation length scale in the vertical direction independently from the horizontal direction. 
More precisely, we incorporate $\gamma_0$ through the function $Q_\phi$, defined as 
\begin{align}\label{eq: Un-scaled Von Karman PSD}
    Q_\phi(\bm{f}; \gamma_0) = \frac{0.023}{\left(f_x^2+\gamma_0 f_y^2+L_0^{-2}\right)^{\frac{11}{6}}}.
\end{align}
Here, $\bm{f}\in\R^2$ is the frequency vector (in cycles per meter) with components $(f_x, f_y)$ in the $x$- and $y$-directions, respectively. The anisotropic Von K\'arman PSD is then 
\begin{align}\label{eq: Von Karman PSD}
    V_\phi(\bm{f}; \gamma_0) = r_0^{-\frac{5}{3} } Q_\phi ( \bm{f}; \gamma_0 ) \ ,
\end{align}
where the standard (isotropic) Von K\'arman PSD is obtained by taking $\gamma_0 = 1$. Equation~(\ref{eq: Von Karman PSD}) shows that the value of $r_0$ determines the scale of the PSD of the phase screens; it thus determines the variance of the phase screens.  We estimate $r_0$ and $\gamma_0$ by fitting Eq.~(\ref{eq: Von Karman PSD}) to the spatial PSD of the measured data as described below.

The estimation of $r_0$ from data requires an estimate of the spatial PSD $\hat{V}_\phi^\text{meas}(\bm{f})$ of the measured data.  
For this we use Welch's method \cite{Welch}, in which we apply a 2D Hamming window to each image $\phi_n$, then find the  average over the magnitude-squared FFTs of each image $\phi_n$ with a 2D Hamming window. 
We then set the resulting estimate $\hat{V}_\phi^\text{meas}(\bm{f})$ equal to the right hand side of Eq.~(\ref{eq: Von Karman PSD}) (with $L_0=\hat{L}_0$) and solve for $r_0$.  Converting from frequency $\bm{f}$ to scaled indices $\bm{k} \Delta_f$, this yields 
\begin{align}\label{eq: r hat Esimation}
    \hat{r}_0(\bm{k}\Delta_f, \gamma_0) 
    = \left(\frac{Q_\phi (\bm{k} \Delta_f; \gamma_0)}{\hat{V}_\phi^\text{meas}(\bm{k}\Delta_f)}\right)^{3/5}.
\end{align}

In the isotropic case, $\hat{r}_0(\bm{k}\Delta_f, \gamma_0=1)$ should be constant, independent of $\bm{k}$, although this is never precisely true for measured data. In the anisotropic case, $\gamma_0 \neq 1$ means that $\hat{r}(\bm{k}\Delta_f, \gamma_0)$ varies with $\bm{k}$. 
Thus, for a fixed $\gamma_0$ we take an average over $\bm{k}$ to estimate $r_0$ as a function of $\gamma_0$: 
\begin{align}\label{eq: r0 estimate}
    \hat{r}_0(\gamma_0) = \text{average}\{\hat{r}(\bm{k}\Delta_f, \gamma_0): \ \bm{k}\in \mathcal{K}\},
\end{align}
where $\mathcal{K}$ is a set of indices $\bm{k}=(k_0, k_1)$ that excludes the indices where $k_0=0$ or $k_1=0$ and excludes the $k_{\max}$ smallest and largest frequency bins along each axis. We exclude the zero indices because the zero frequencies are distorted by TTP removal, and we exclude the edge frequencies because they can be contaminated by aliasing. The frequency cut-off $k_{\max}$ is data-dependent.

Finally, we estimate $\gamma_0$ by fitting to the 2D structure function of the measured data, which is defined in Sec.~\ref{s: Two-Dimensional Phase Structure Function}. The level sets of the 2D structure function are elliptical, with eccentricity varying with $\gamma_0$.  Each value of $\gamma_0$ yields an estimate $\hat{r}_0(\gamma_0)$, which together yield a corresponding 2D structure function.  Hence, we do a 1D optimization in $\gamma_0$ to minimize the mean-squared error between the structure function for this $\gamma_0$ and the structure function of the measured data.  We take $\hat{\gamma}_0$ to be the value that minimizes this error.
If the measured data is assumed to be isotropic, we choose $\hat{\gamma}_0 = 1$.

\subsection{Estimating the Flow Velocity}
\label{s: Estimation of Temporal Parameters}
The velocity vector $\bm{v} = (v_x, v_y)$ gives the translation (or flow) in pixels from time step $n-1$ to $n$ as in Eq.~(\ref{eq: Flow}) (both assumed to be TTP-removed). After $T$ time steps, we obtain a pixel shift of $T\bm{v}$, which gives rise to a correlation peak between pixels $\phi_{n-T}(\bm{i}-T\bm{v})$ and $\phi_n(\bm{i})$ where $\bm{i} = (i_x, i_y)$ and we use interpolation for fractional pixel shifts. Thus, we can estimate $\bm{v}$ by locating correlation peaks between $\phi_n$ and the $T$-step flow of $\phi_{n-T}$. However, choosing $T$ is difficult because (1) small values of $T$ may not give precise estimates due to numerical ill-conditioning and (2) large values of $T$ will include frames with significant boiling, leading to de-correlation with the earlier frames. To mitigate this issue, we average the estimates over multiple values of $T$.

\begin{figure}
    \centering
    \includegraphics[width=0.85\linewidth]{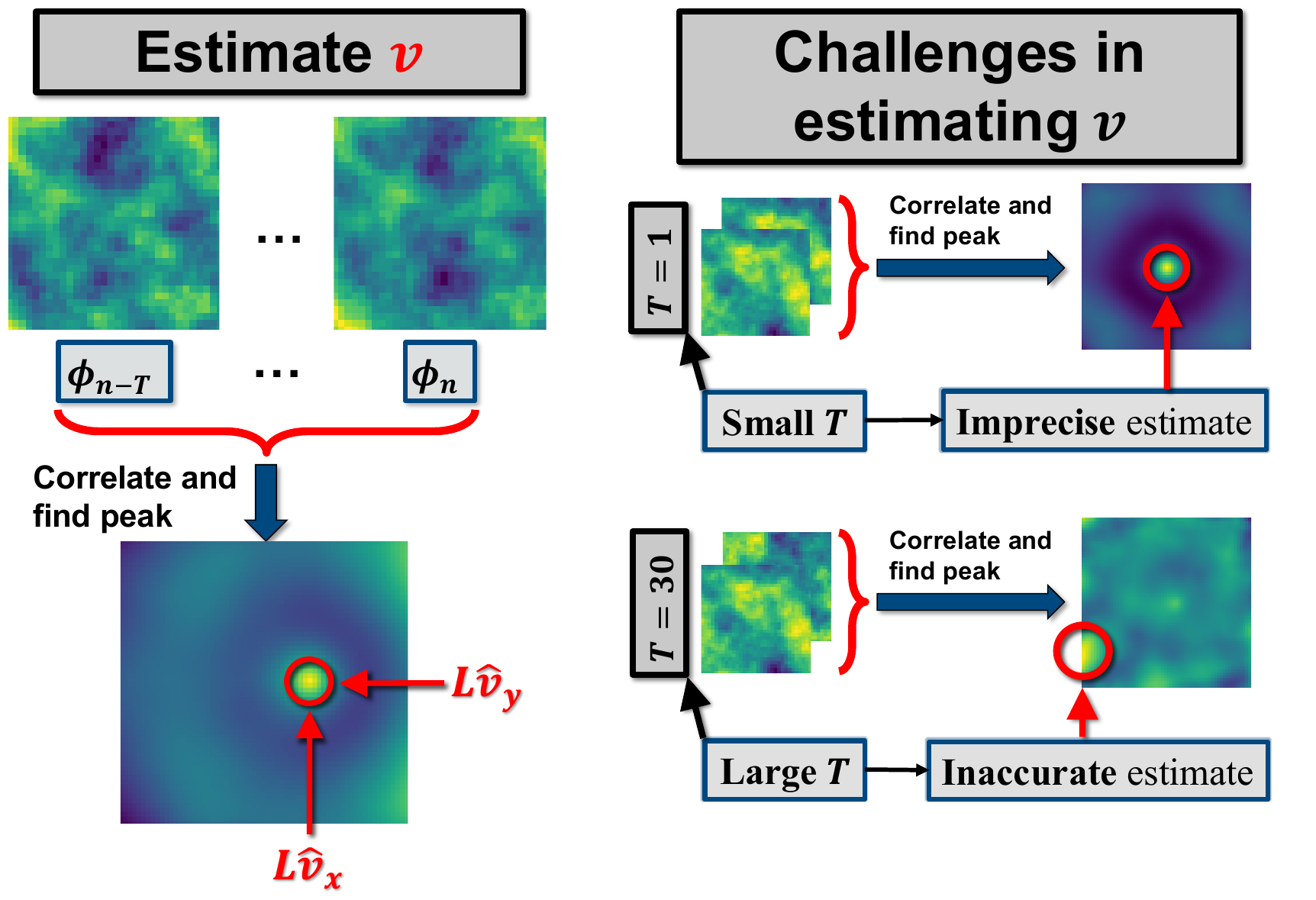} \hfill

    \caption{{\bf Left:} Estimation of the flow velocity $\bm{v}$. Given a time-lag $T$, we find $\bm{v}$ to maximize the correlation between $\phi_n$ and Flow$(\phi_{n-T}; T \bm{v})$, then average $\bm{v}(T)$ over $T$ between 1 and $T_{\max}$. {\bf Right:} Uncertainty in the time-shift $T$. Small values of $T$ yield imprecise estimates of the flow velocity due to limited data and numerical ill-conditioning, while large $T$ yield inaccurate estimates due to the decorrelation induced by boiling.}
    \label{fig: Flow Velocity Estimation}
\end{figure}

Figure~\ref{fig: Flow Velocity Estimation}, left, illustrates our estimation algorithm for $\bm{v}$. For each time-shift $T$, we find the cross-correlation between frames at $n-T$ and $n$ as a function of $\bm{v}$ as 
\begin{align}\label{eq: Cross-Correlation}
    R_{\phi'}(\bm{v}; T) = \E\left[\sum_{\bm{i}} \phi_{n-T}'(\bm{i} - T\bm{v} ) \; \phi'_n(\bm{i})\right],
\end{align}
where $\phi_n'$ denotes the mean-subtracted $\phi_n$ divided by its standard deviation, and we approximate the expected value using an average over $n$ (which we denote $\hat{R}_{\phi'}$).   We then take a maximum over $\bm{v}$ to find
\begin{align}\label{eq: Max Cross-Correlation}
    \hat{\bv}(T) = \frac{1}{T}\times \underset{\bv}{\text{argmax}}\;\hat{R}_{\phi'}(\bv;T).
\end{align}
Finally, we average $\hat{v}(T)$ over $T \in \{1, \ldots, T_{\max}-1\}$, where $T_{\max}$ is chosen based on conditions we describe next. 

Figure~\ref{fig: Flow Velocity Estimation}, right, illustrates the considerations in choosing the value of $T_{\max}$. Small values of $T_{\max}$ may not yield a precise estimate due to limited data and numerical ill-conditioning. Conversely, large values of $T_{\max}$ may give inaccurate estimates because the cross-correlation $R_{\phi'}$ becomes contaminated by boiling. We thus determine $T_{\max}$ based on two criteria. First, note that for sufficiently large values of $T$ depending on the true value of $\bm{v}$, the shift $\bm{i} - T\bm{v}$ will lie outside the aperture for all $\bm{i}$. In this case, the estimate $\hat{R}_{\phi'}(\bm{v};T)$ will not capture the correlations induced by flow. Thus, we constrain
\begin{align}\label{eq: T_max 1}
    T_{\max} \leq \Bigg\lfloor \frac{N-1}{\|\hat{\bv}\|_{\max}}\Bigg\rfloor,
\end{align}
where $\|\hat{\bv}\|_{\max}$ is the largest magnitude $\|\hat{\bv}(T)\|$ over all time-lags $T$, and $N$ is the side length in pixels.
Additionally, for large $T$ depending on the rate of boiling, the boiling component may obscure the correlations due to flow.  To prevent this, we require that the signal-to-noise ratio (SNR) of the estimate $\hat{R}_{\phi'}(\bm{v}; T)$ be at least $10$ for each $T$. The resulting bound for $T_{\max}$ is given in Appendix~\ref{appendix: Tolerance}. 

\subsection{Estimating the Flow-Coefficient}

\begin{figure}
    \centering
    \includegraphics[width=0.6\linewidth]{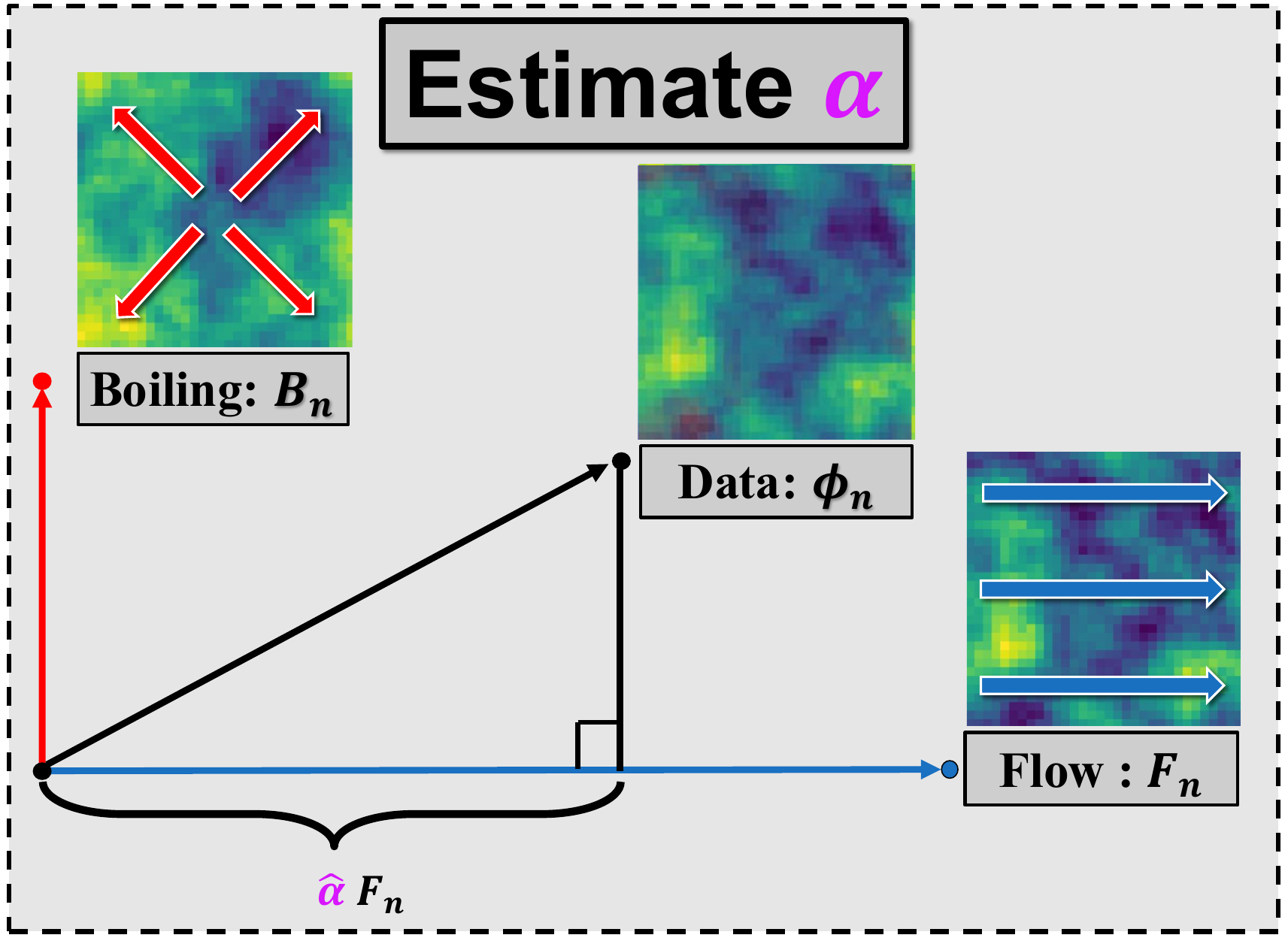} 
    \caption{Estimation of the flow-coefficient $\alpha$. The expected inner product between boiling $B_n$ and flow $F_n$ is 0, so we estimate the flow-coefficient $\alpha$ by projecting the data onto the 1D span of $F_n$.  In practice, we do this projection in Fourier space since $F_n$ is computed in Fourier space, and we make use of all times steps $n$.}
    \label{fig: alpha estimation}
\end{figure}

The flow-coefficient $\alpha$ determines the effect of flow relative to boiling.  
From Eq.~\eqref{eq: Boiling Flow}, $\phi_n = \sqrt{1-\alpha^2}\:B_n + \alpha \:F_n$ is a linear combination of boiling $B_n$ and the flow $F_n$ of $\phi_{n-1}$. 
Since $B_n$ and $\phi_{n-1}$ are zero-mean Gaussians which are independent of one another, the expected inner product $\E[\langle B_n, F_n \rangle]$ is 0.  
Hence we regard $B_n$ and $F_n$ as orthogonal vectors and project $\phi_n$ onto the span of $F_n$ to estimate $\alpha$.
This estimation of $\alpha$ is equivalent to finding $\alpha$ to minimize the norm of $\phi_n - \alpha F_n$; we use this latter formulation since it is more amenable to incorporating all times step simultaneously.  We also do the calculation in Fourier space since the flow is performed natively in Fourier space.  

Figure~\ref{fig: alpha estimation} illustrates the geometric intuition for estimating $\alpha$. In practice, we first estimate $\bm{v}$ as above, then compute $\tilde{F}_n$ in Eq.~(\ref{eq: Flow}) using FFTs of the measured data (assumed to be TTP-removed as above). In addition, we remove TTP from $\tilde{F}_n$ and enforce conjugate-symmetry. We then find the least-squares fit in Fourier space:
\begin{align}\label{eq: Boiling Flow alpha Problem}
    \hat{\alpha} = \underset{\alpha}{\text{argmin}}\left\{\sum_{n=1}^{N_T-1}\left\|\tilde{\phi}_n - \alpha \;\tilde{F}_n \right\|^2\right\}.
\end{align}
Here, $N_T$ is the number of time-steps of training data. As with estimating $r_0$, we exclude any frequency bins containing a zero frequency and the smallest and largest $k_{\max}$ frequency bins along each axis to reduce the effects of TTP-removal and potential aliasing. Taking the derivative with respect to $\alpha$ allows us to find $\hat{\alpha}$ using only two inner products.

\section{Two-Dimensional Structure Function}\label{s: Two-Dimensional Phase Structure Function}
In this section, we define the two-dimensional phase structure function. Given a separation vector $\br \in \R^2$, the 2D phase structure function is given by
\begin{align}\label{eq: 2D Phase Structure Function}
    D_\phi(\br) = \E\left[\left(\phi(\x+\br) - \phi(\x)\right)^2\right],
\end{align}
where $\phi(\x)$ is the phase error at grid location $\x = (x,y)$, and the expectation is over phase screens $\phi$. We make the assumption that $\phi$ is spatially stationary, so that $D_\phi$ is not a function of $\x$. The phase structure function has been widely used to characterize the phase aberrations imposed by atmospheric turbulence \cite{Vogel, Schmidt, Siegenthaler}.

In the case of atmospheric turbulence, the phase structure function depends only on the magnitude of $\br$. Specifically, because phase errors induced by atmospheric turbulence are spatially homogeneous and isotropic \cite{Vogel, Siegenthaler, Schmidt}, Eq.~(\ref{eq: 2D Phase Structure Function}) follows the Kolmogorov fifth-third power law 
\begin{align}\label{eq: Kolmogorov Power Law}
    D_\phi(\br) = 6.88 \left(\frac{\| \br\|}{r_0}\right)^{5/3},
\end{align}
where $r_0$ is the Fried coherence length. This property is instrumental for generating atmospheric phase screens \cite{Schmidt}. Furthermore, because Eq.~(\ref{eq: Kolmogorov Power Law}) is independent of the angle $\angle \br$, it is inherently a function of the one-dimensional separation distance $\|\br\|$. 

However, for anisotropic aero-optic effects, $D_\phi$ depends on the angle of $\bm{r}$ and the magnitude\cite{Rasouli, Silbaugh}, so Eq.~(\ref{eq: Kolmogorov Power Law}) no longer applies.
Thus, we compute Eq.~(\ref{eq: 2D Phase Structure Function}) as a function of the two-dimensional separation vector $\br$.

\subsection{Estimation from Discrete Data}\label{s: Estimation From Discrete Data}
Although $\br$ is often presented as a continuous input with units of distance, we estimate the structure function from discrete data $\phi_n\left(\bm{i}\right)$, where $\bm{i} = (i_x, i_y)$ lies in a discrete set of indices into the 2D array $\phi_n$.
To avoid interpolation, we restrict $\bm{r}$ to vectors of the form $(\bm{i}_1 - \bm{i}_2) \Delta $, where $\Delta>0$ is the pixel width in meters.

To estimate the structure function, we first normalize a phase screen as in the estimation of $\bm{v}$ to obtain $\phi'$.  
Using the approach of Ref.~\citenum{Vogel}, we compute a quasi-homogeneous structure matrix
\begin{align}\label{eq: Quasi-Homogeneous Structure Function}
   \hat{D}_{\phi'}^{qh}\left(\bm{i}_1, \bm{i}_1\right) = 2\left(1 - \E\left[\phi'(\bm{i}_1) \; \phi'(\bm{i}_2)\right]\right),
\end{align}
where we use time-averages to approximate the expected value. 
Since the expectation in Eq.~\eqref{eq: 2D Phase Structure Function} is unchanged after replacing $\bm{r}$ with $-\bm{r}$, 
we estimate the 2D phase structure function via
\begin{align}\label{eq: Estimate of Two-Dimensional Function}
    \hat{D}_{\phi}(\br) = \text{average}\left\{\hat{D}_{\phi'}^{qh}\left(\bm{i}_1, \bm{i}_2\right):\hspace{0.3cm} \bm{i}_1, \bm{i}_2 \text{ satisfy }  (\bm{i}_1-\bm{i}_2) \Delta = \pm \br \right\}.
\end{align}

\section{Data and Metrics}\label{s: Data and Metrics}
In this section, we describe the simulated and measured data along with the metrics used to evaluate our parameter estimation methods.

\subsection{Simulated Data}\label{s: Simulated data sets}
We generate simulated data by choosing boiling flow parameters and then generating data using Eqs.~\eqref{eq: Boiling}, \eqref{eq: Flow}, and \eqref{eq: Boiling Flow}.
For each data set, we fix the outer scale $L_0$ (equal to the aperture length), Fried coherence length $r_0$, anisotropy parameter $\gamma_0=1$, vertical flow velocity component $v_y=0$, and temporal sampling rate $f_s = 100$ kHz. 

\begin{table}[htbp]
    \centering
    \caption{Ground-Truth Parameters of Isotropic Simulated Data.  Note that each line for $v_x$ corresponds to physical flow velocities $17.2, 34.4, 68.8, 137.5$ in meters per second.}
    \label{tab: Simulated Data Parameters}
    \begin{tabular}{|c|c|c|c|c|}
        \hline

        \multicolumn{5}{|c|}{\centering {\bf Fixed Parameters}} \\
        \hline
        \hline 
        Aperture Width & Fried Param. &  Anisotropy Param. & Vertical Vel. & Temp. sampling rate\\
        \hline
        $L_0=44$ mm & $r_0=4.4$ mm  & $\gamma_0=1$ & $v_y=0$ & $f_s = 100$ kHz\\
        \hline
        \addlinespace[0.5em]
    \end{tabular}

    \vspace{12pt}
    \begin{tabular}{|c|c|c|}
        \hline

        \multicolumn{3}{|c|}{\centering {\bf Variable Parameters}} \\
        \hline        
        \hline
        Image Dimension N & Flow Velocity $v_x$ & Flow-Coefficient $\alpha$ \\
        (pixels) & (pixels per time-step) & (unitless)\\
        \hline
        20 & 0.08, 0.16, 0.31, 0.63 & \multirow{3}*{$0.1, 0.3, 0.5, 0.7, 0.9$}\\
        32 & 0.13, 0.25, 0.5, 1.0 & \\
        64 & 0.25, 0.5, 1.0, 2.0 & \\
        \hline 
    \end{tabular}
\end{table}

We investigate the effects of pixel width, flow velocity, and flow-coefficient by varying $N$, $v_x$, and $\alpha$ for different data sets. 
We use physical flow rates of $17.2, 34.4, 68.8, 137.5$ meters per second and convert to pixels per time step as follows.  
With a fixed aperture width of $L_0$ meters, the pixel width is given by $\Delta = L_0 / N$ in meters per pixel, and so $v_x = \text{(velocity in m/s)} / (f_s \Delta)$ in pixels per time step.

Table~\ref{tab: Simulated Data Parameters} shows the parameter values used to generate simulated data.

\subsection{Measured Data}
\label{s: Measured data sets}
To evaluate results from measured data, we used two turbulent boundary layer (TBL) data sets containing measured aero-optic phase aberrations obtained from a high-speed wind tunnel experiment\cite{Kemnetz, KemnetzDissertation}. The phase screen data was measured using a Shack Hartmann Wavefront Sensor, with TTP removed from the data.

Table~\ref{tab: Experimental Data Sets} shows details of the resulting aero-optic phase screen data. Note that the image dimensions $M\times N$ are not square.  
Additionally, a subset of pixels near the boundary had no data.  
Therefore, we generated boiling flow data with dimension $\max\{M, N\}$ and applied a mask to match the largest inscribed square containing valid data. 
Since the measured data showed contamination at low temporal frequencies (i.e., sharp peaks at frequencies below 1 kHz, possibly due to vibrations), we pre-processed both measured data sets to remove these additional signals; Appendix~\ref{appendix: Pre-Processing} describes this process.

\begin{table}[htbp]
    \centering
    \caption{Measured Data Sets F06 and F12}
    \begin{tabular}{|c||c|c|}
        \hline 
        & F06 & F12 \\
        \hline
        Image Dimension $M \times N$ & $25\times 34$ & $21\times 22$ \\
        Largest Inscribed Square Length [pixels] & 22 & 18 \\
        Pixel Spacing $\Delta$ [mm] & 1.43 & 2.24 \\
        Sampling Frequency $f_s$ [kHz] & 100 & 130 \\
        \hline 
    \end{tabular}
    \label{tab: Experimental Data Sets}
\end{table}

\subsection{Quality Metrics}\label{s: Quality Metrics}
We evaluate our parameter fitting method using the temporal power spectrum (TPS) and the 2D structure function. We compute the TPS of the phase screens themselves and of the spatial finite difference of the phase screens in the $x$-direction (i.e., the deflection angle $\theta_x$ \cite{Kemnetz, SiegenthalerShear, Cress}).  For a phase screen $\phi$, we call these $S_\phi$ and $S_{\theta_x}$, respectively.  We estimate the TPS of each pixel (in units of energy per second) using the approach of Refs.~\citenum{PoyneerExperimental, Oppenheim} by applying a Hamming window on multiple subsets of the time samples and then averaging the resulting TPS estimates over all pixels and subsets.
We compute the 2D structure function $D_\phi$ using the approach outlined in Sec.~\ref{s: Estimation From Discrete Data}.

To evaluate the accuracy of parameter estimation for $v_x$ and $\alpha$, we plot the relative errors $\|\hat{\bm{v}}-\bm{v}\| / \|\bm{v}\|$ and $|\hat{\alpha}-\alpha| / |\alpha|$, and we report the relative error for $r_0$.

To measure the statistics of generated data compared to given data, we use normalized root-mean squared error (NRMSE). Here, the normalization is given by dividing by $P_{95}(\bm{y}) - P_5(\bm{y})$, where $P_{m}(\bm{y})$ denotes the $m$th percentile of data $\bm{y}$. Specifically, given two arrays $\bm{y}, \bm{y}^{\text{data}}\in\R^K$, we define
\begin{align}\label{eq: NRMSE}
    \text{NRMSE}(\bm{y}, \bm{y}^{\text{data}}) = \frac{\frac{1}{\sqrt{K}}\bigl\|\bm{y}-\bm{y}^{\text{data}}\bigr\|_2}{P_{95}(\bm{y}^{\text{data}})-P_5(\bm{y}^{\text{data}})},
\end{align}
where $\|\cdot\|_2$ is the vector $L_2$ norm. 

We report the following errors, where in each case a superscript ``data'' indicates values obtained using the simulated or measured data, and a hat indicates values obtained from generated data using estimated parameters.
\begin{itemize}
    \item \textbf{Flow TPS Error}: $\text{NRMSE}(\hat{S}_{\theta_x}, S_{\theta_x}^{\text{data}})$. 

    \item \textbf{Phase TPS Error}: $\text{NRMSE}(\hat{S}_\phi, S_\phi^{\text{data}})$.

    \item \textbf{Structure Function Error}: $\text{NRMSE}\left(\sqrt{\hat{D}_{\phi}}, \sqrt{D_{\phi}^{\text{data}}}\right)$.  
\end{itemize}
We take the square root of $D_{\phi}$ to more evenly weight the large and small values of the structure function.  

\section{Results}
\label{s: Results}
In this section, we show results from the parameter estimation algorithm described in Sec.~\ref{s: Parameter Estimation from Training Data} applied to both isotropic simulated data and anisotropic measured aero-optic data. 
For all experiments, we used the frequency bin cut-off $k_{\max}=2$ when estimating $r_0$ and $\alpha$.
We evaluate our method using the error metrics of the previous section.

\subsection{Results from Simulated Data}\label{s: Results from simulated data}
For each choice of ground-truth parameters, we generated ten independent, $10K$ time-step `ground-truth' boiling flow data sets for parameter estimation and an additional $50K$ time-step data set for error evaluation.  For each of these ten generated data sets, we applied our estimation algorithm for isotropic phase screens (i.e., setting $\hat{\gamma}_0=1$) and then used the resulting parameter estimates to generate $50K$-time-step boiling flow data sets. We computed the errors from Sec.~\ref{s: Quality Metrics} using this generated $50K$ data set along with the $50K$ evaluation data, then averaged over the ten data sets. 

\subsubsection{Parameter estimation results}\label{s: Parameter estimation results}
\begin{figure}[t]
    \centering
    \includegraphics[width=0.8\textwidth]{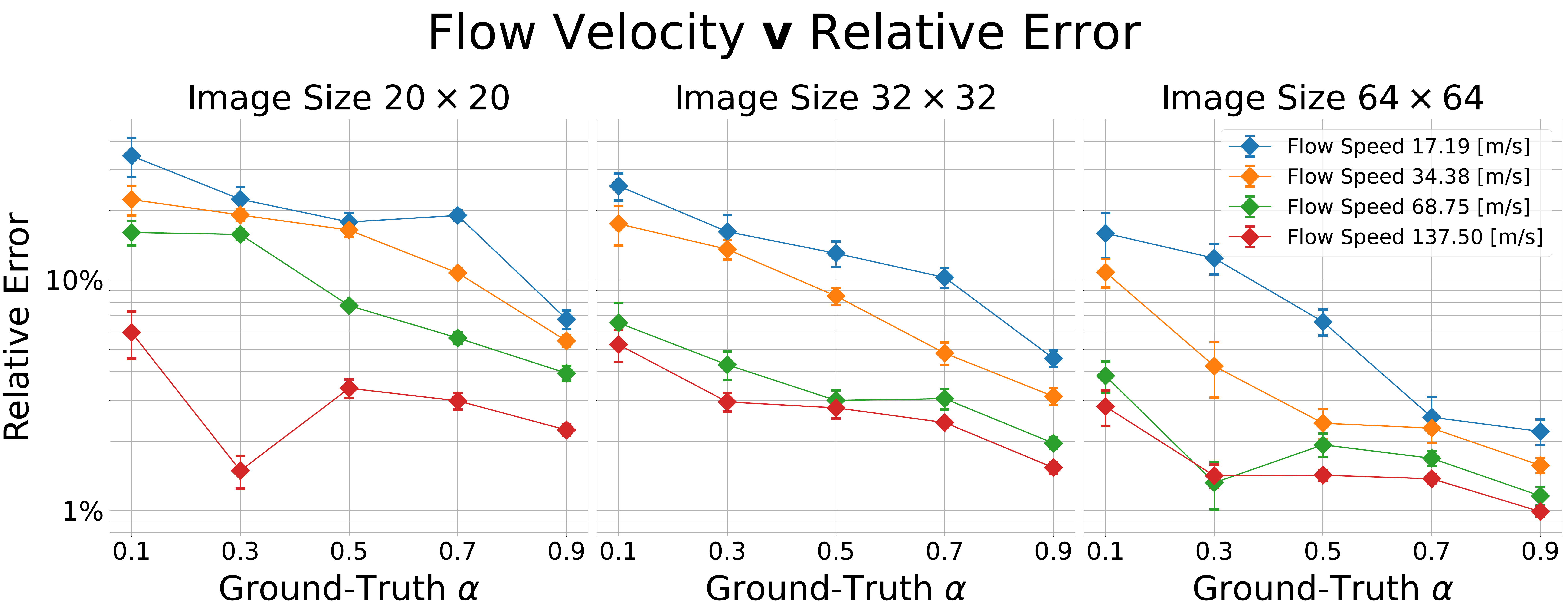}
    \includegraphics[width=0.8\textwidth]{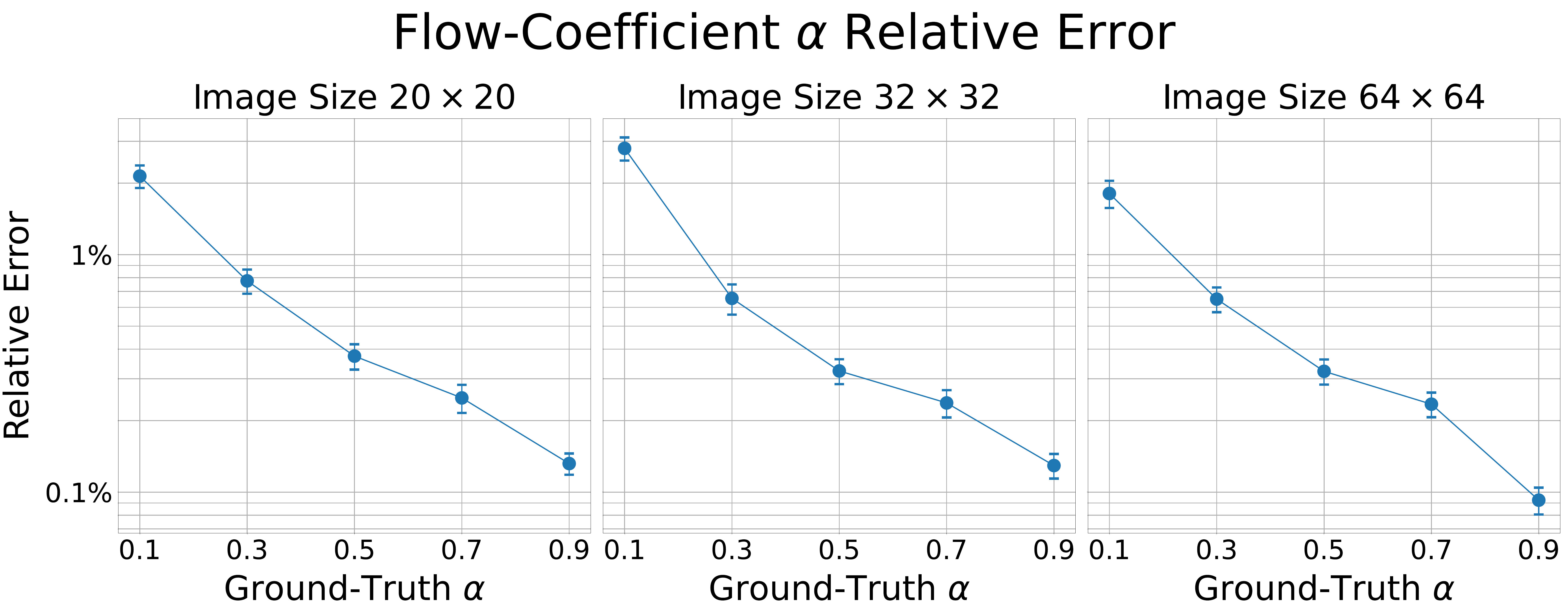}
    \caption{{\bf Top:} Relative errors of the flow velocity estimates for various image sizes, ground-truth values of $\alpha$ (horizontal axis), and physical velocity (color). Note that the errors are generally smallest for (1) large ground-truth values of $\alpha$ (less boiling relative to flow), (2) large ground-truth flow speeds (larger pixel displacements per time step), and (3) large images. The errors are below 10\% in the majority of cases.  {\bf Bottom:} Relative errors of the flow-coefficient estimates $\hat{\alpha}$ for various image sizes and ground-truth values of $\alpha$. Note that the errors are smallest for large ground-truth values of $\alpha$ (less boiling relative to flow) and are slightly lower for larger image sizes. Importantly, the errors remain below 4\% in all cases.}
    \label{fig: Flow Velocity Error}
\end{figure}

Figure~\ref{fig: Flow Velocity Error}, top, plots the relative error $\|\hat{\bv}-\bv\| / \|\bv\|$ of the flow velocity estimates as a function of various ground-truth values of $v_x$ and $\alpha$.  Our velocity estimates are generally more accurate for larger ground-truth values of $\alpha$ (less boiling relative to flow) and larger ground-truth flow velocities (larger pixel displacements per time step); there is also a small improvement in the estimates for larger images. Although the errors can exceed 30\% in the edge case that $N=20$ and the ground-truth $\alpha$ is 0.1, the errors remain below 10\% in the majority of cases.

Figure~\ref{fig: Flow Velocity Error}, bottom, plots the relative error $|\hat{\alpha} - \alpha| / |\alpha|$ of the flow-coefficient estimates $\hat{\alpha}$ as a function of the ground-truth $\alpha$. 
This error showed little dependence on ground-truth flow speeds, so we averaged the errors over all ground-truth $v_x$. The estimates $\hat{\alpha}$ are significantly more accurate for larger ground-truth values of $\alpha$. However, there is only a marginal improvement in accuracy for larger images. The errors remain below 1\% in most cases and rise to 2-3\% only when the ground-truth $\alpha$ is 0.1.

The accuracy of the estimate $\hat{r}_0$ depends only on the image size. Different ground truth velocities and flow-coefficients $\alpha$ do not modify the spatial PSD estimate of data with sufficiently many time-steps, so the estimate $\hat{r}_0$ does not vary significantly across different ground-truth values of $v_x$ and $\alpha$. The average relative errors of the $r_0$ estimates were 3.8\%, 1.9\%, and 1.1\% for $N=20$, $32$, and $64$, respectively. Thus, the accuracy of the estimates increases significantly with larger image sizes but the errors remain below 4\% in all cases.

\subsubsection{Simulated data TPS and structure function errors}\label{s: Generated data results}

\begin{figure}[b]
\centering
    \includegraphics[width=0.8\textwidth]{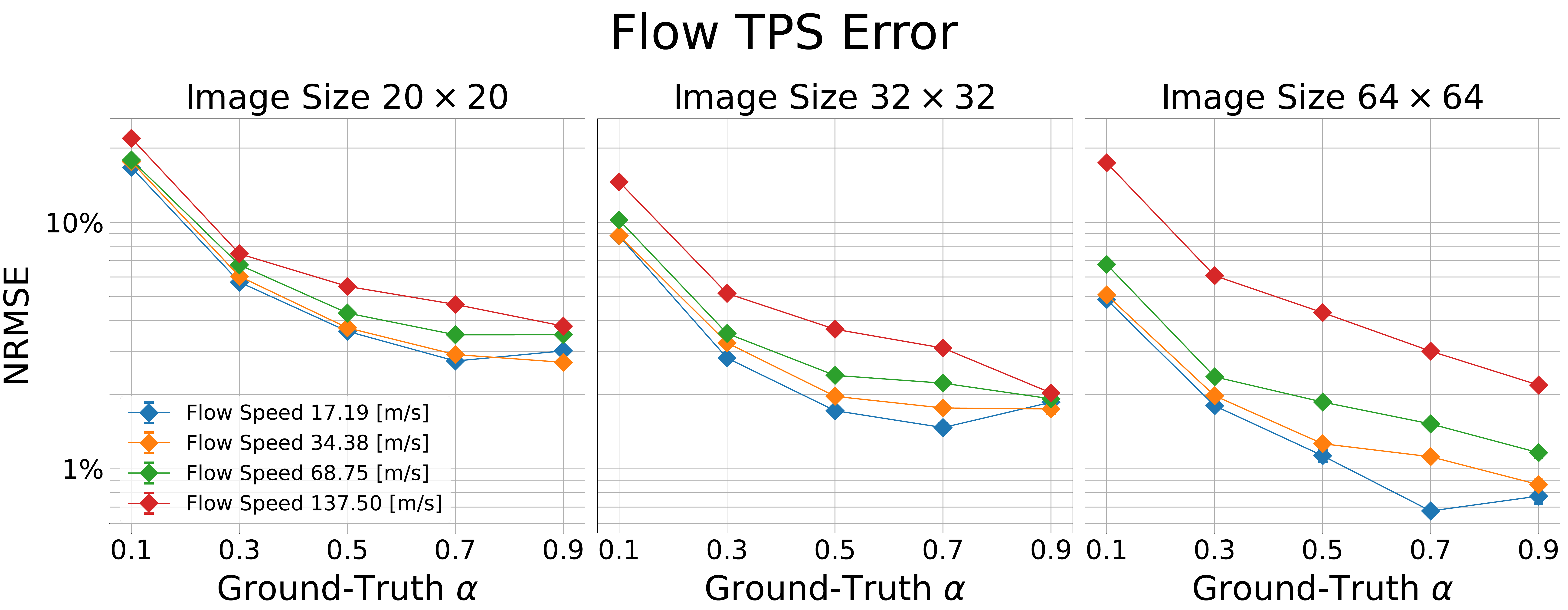}
    \includegraphics[width=0.8\textwidth]{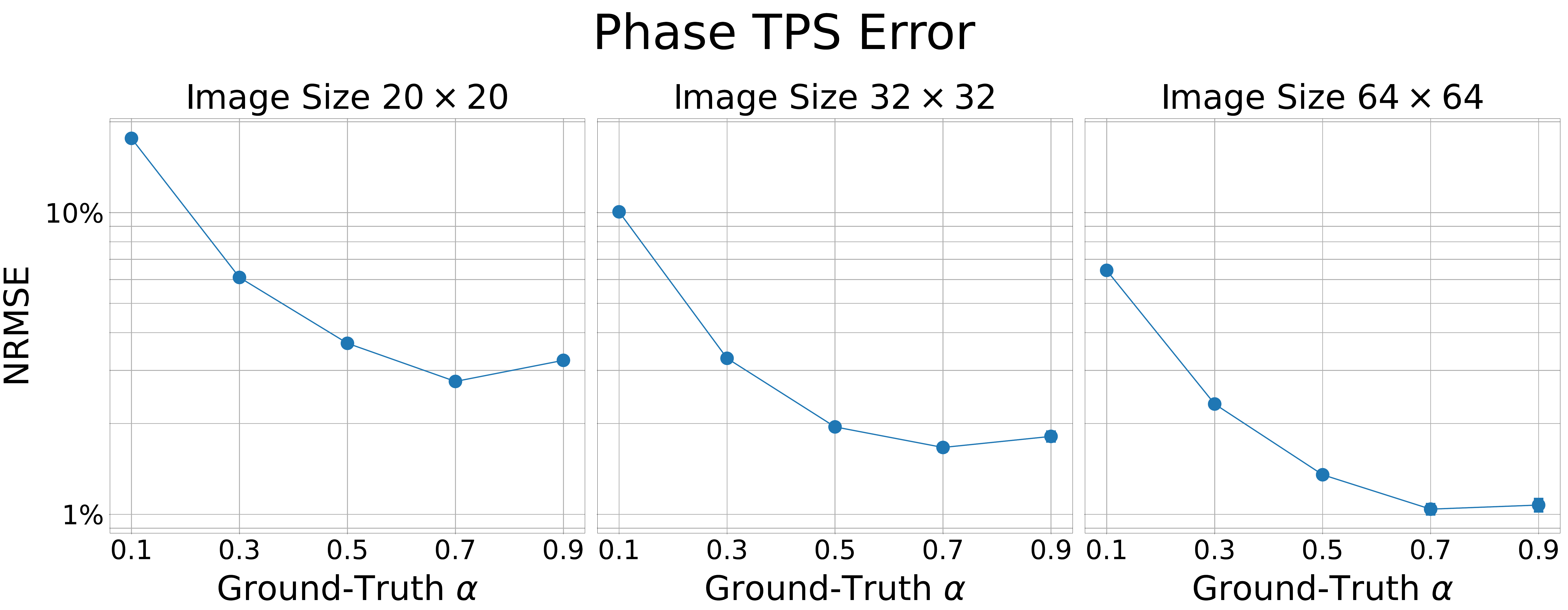}
    \caption{{\bf Top:} Flow Temporal Power Spectrum (TPS) NRMSE for various image sizes, ground-truth values of $\alpha$ (horizontal axis), and ground truth velocity (color). Note that the errors are smallest for (1) larger images, (2) larger ground-truth values of $\alpha$, and (3) smaller ground-truth velocities. While the errors can exceed 20\% when $\alpha=0.1$, they remain below 10\% when $\alpha >0.1$. 
    {\bf Bottom:} Phase Temporal Power Spectrum (TPS) NRMSE for various ground-truth values of $\alpha$. The errors are smallest for larger images and larger ground-truth values of $\alpha$, and they exceed 10\% only in the edge case with $20\times 20$ images and $\alpha=0.1$.}
    \label{fig: Flow TPS Error Plot}
\end{figure}

Figure~\ref{fig: Flow TPS Error Plot}, top, plots the flow TPS error for various ground-truth values of $\alpha$ and physical velocity. Consistent with the estimates of $\hat{r}_0$ and $\hat{\alpha}$, the flow TPS errors decrease significantly with larger images and larger ground-truth values of $\alpha$. However, since the flow TPS is sensitive to changes in the flow velocity, small errors in the estimate $\hat{\bv}$ can cause large flow TPS errors. As a result, even though the relative error in $\hat{\bv}$ decreases with larger ground truth velocities, the absolute error increases, and so the flow TPS errors also increase with larger flow speeds. Despite this, the errors exceed 10\% only when the ground-truth $\alpha$ is $0.1$, and they remain below 8\% when $\alpha > 0.1$.

Figure~\ref{fig: Flow TPS Error Plot}, bottom, plots the phase TPS errors for various ground-truth values of $\alpha$. Unlike the flow TPS, the phase TPS has only weak dependence on the flow velocity, so we average over velocity and plot as a function of $\alpha$ alone. As with the errors of the estimates $\hat{r}_0$ and $\hat{\alpha}$, the phase TPS error decreases for larger images and larger ground-truth values of $\alpha$. Specifically, the errors exceed 10\% only in the edge case that $N=20$ and the ground-truth $\alpha$ is 0.1.

The structure function NRMSE defined in Sec.~\ref{s: Quality Metrics} depends at most weakly on the ground-truth values of $\alpha$ or velocity because the estimate~(\ref{eq: Estimate of Two-Dimensional Function}) averages over time and boiling flow generates temporally stationary data. Averaging over everything except image size, the average structure function NRMSE values were below 1.2\% for each image size, with the error varying only marginally as a function of image size.

\subsubsection{Simulated data TPS and structure function examples}\label{s: tps example}
Figure~\ref{fig: TPS Results 64, v_x137.5, alpha0.9} shows the TPS and structure function results for simulated data with $N=64$, ground truth velocity of $137.5$ (m/s), and $\alpha=0.9$. With this choice of parameters, the estimated TPS and structure function both closely match the reference TPS and structure function.

\begin{figure}[h]
    \centering
    \includegraphics[width=0.7\textwidth]{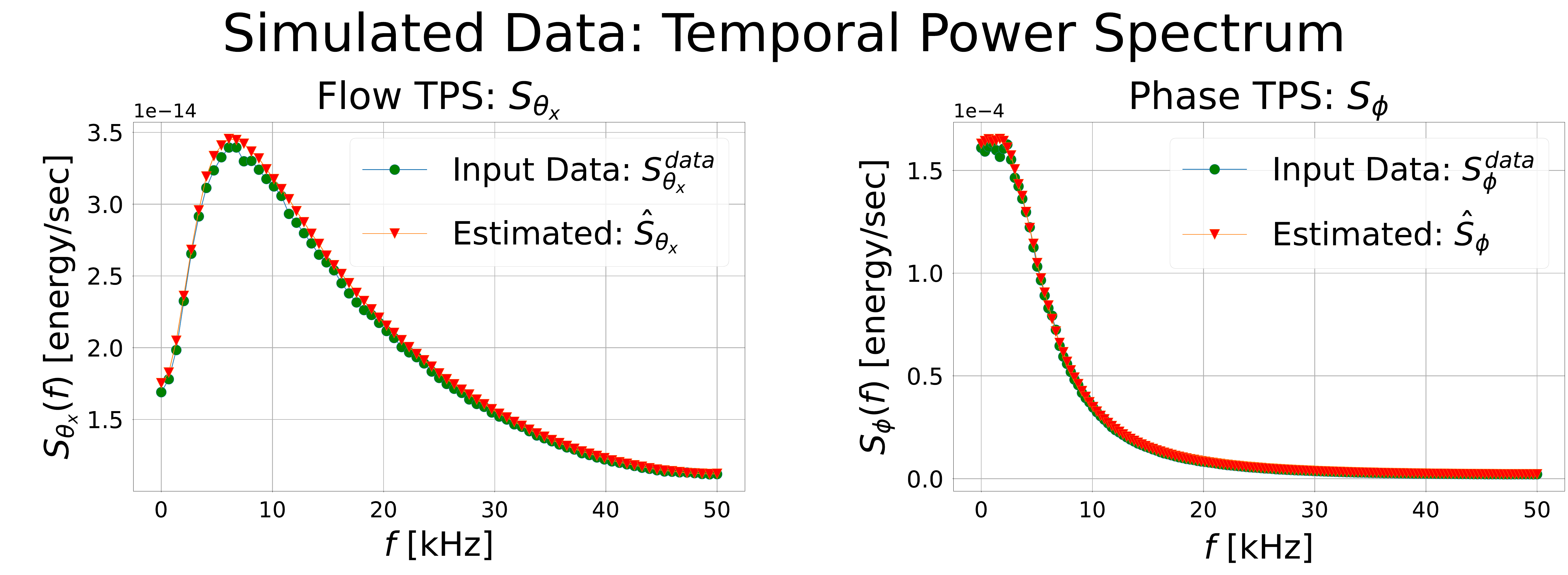}
    \includegraphics[width=0.6\textwidth]{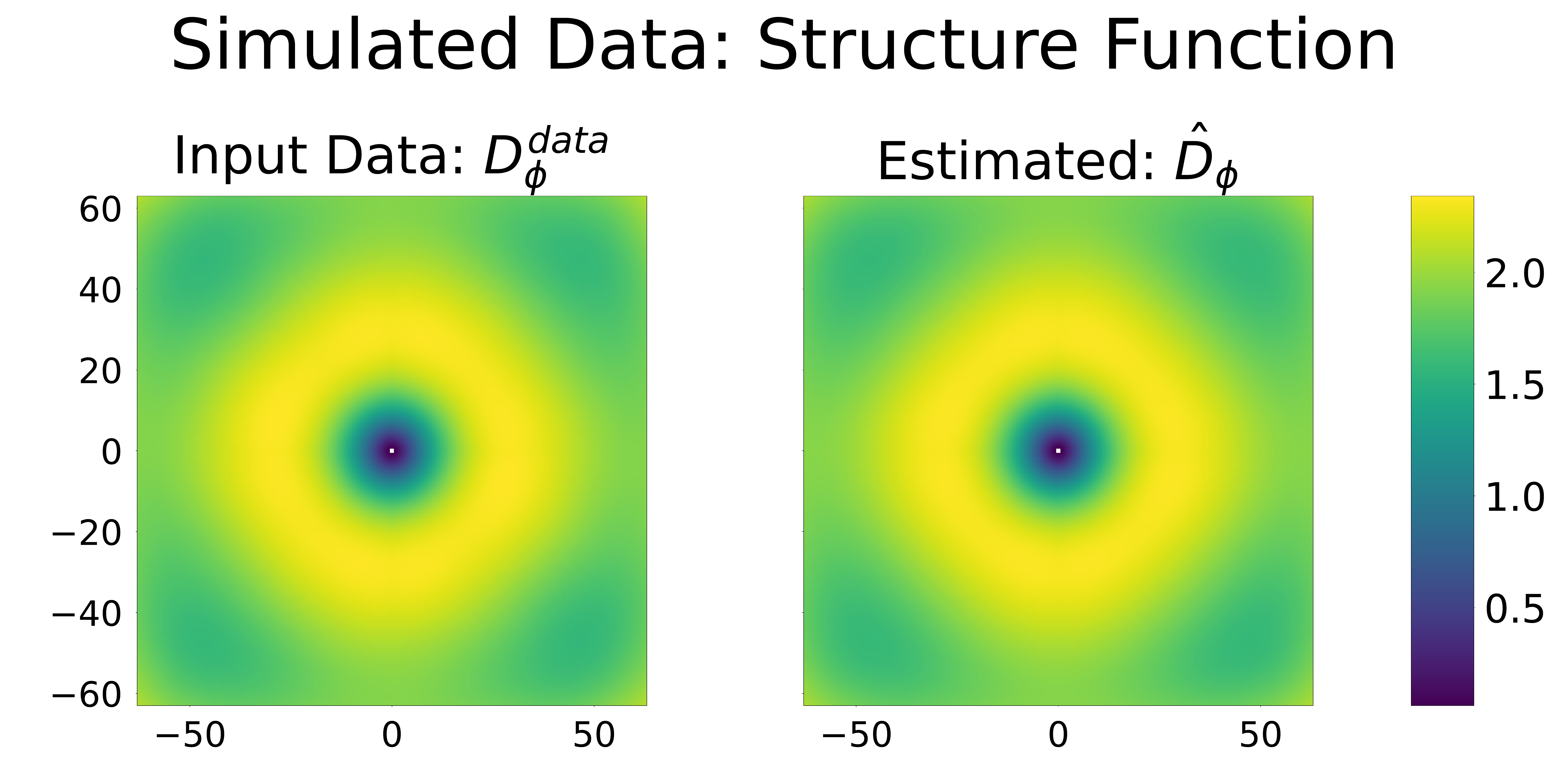}
    \caption{Isotropic estimation from simulated data. {\bf Top:} Comparisons of power spectra obtained from input data (blue) and from isotropic boiling flow data using estimated parameters (orange). The input data has image size $64\times 64$, flow speed $v_x=137.5$ (m/s), and $\alpha=0.9$.   
    The estimated TPS closely matches the reference TPS for both flow, $S_{\theta_x}$, and phase, $S_\phi$.  
    {\bf Bottom:} Comparison of the structure functions~(\ref{eq: Estimate of Two-Dimensional Function}) obtained from input data (left) and data using estimated parameters (right). The image size, flow speed, and $\alpha$ are the same as the top figure. The estimated structure function closely matches the reference structure function.}
    \label{fig: TPS Results 64, v_x137.5, alpha0.9}
\end{figure}

\subsection{Results from Measured Data}
\label{s: Results from measured data}
We applied our parameter estimation algorithm to each data set F06 and F12 from Table~\ref{tab: Experimental Data Sets}. We used the first 80\% of each measured time series for estimation and used the remaining 20\% for error evaluation.  For each data set, we estimated boiling flow parameters using both standard isotropic boiling flow ($\hat{\gamma}_0=1$) and our newly introduced anisotropic boiling flow ($\hat{\gamma}_0$ estimated from data).  We then used these parameter estimates to generate ten isotropic and anisotropic data sets, each with ten times the samples of the evaluation time series. We computed the errors from Sec.~\ref{s: Quality Metrics} using these generated data sets versus the 20\% error evaluation data, then averaged over the ten data sets.

\subsubsection{Isotropic estimation}
Table~\ref{tab: Isotropic Parameter Estimates} shows the parameter estimates obtained using isotropic boiling flow to fit each measured data set. The large values of $\hat{\alpha}$ indicate that the measured data sets are dominated by flow and have minimal boiling. Further, the flow velocity estimates $\hat{\bv}$ show that each flow is streamwise along the $x$-axis and varies by a factor of 2 between data sets F06 and F12.

\begin{table}[htbp]
    \centering
    \caption{Parameter Estimates from Measured Data - Isotropic Boiling Flow}
    \label{tab: Isotropic Parameter Estimates}
    \begin{tabular}{|c||c|c|c|c|}
        \hline
        \diagbox{Data Set}{Parameter} & $\hat{L}_0$ [mm] & $\hat{r}_0$ [mm] & $\hat{\bv}$ [pixels per time-step] & $\hat{\alpha}$ \\
        \hline 
        F06 & 48.62 & 144.26 & (1.20, -0.02) & 0.96 \\
        F12 & 49.28 & 141.78 & (0.55, -0.01) & 0.95 \\
        \hline
    \end{tabular}
\end{table}

\begin{figure}[t]
    \centering
    \includegraphics[width=0.7\textwidth]{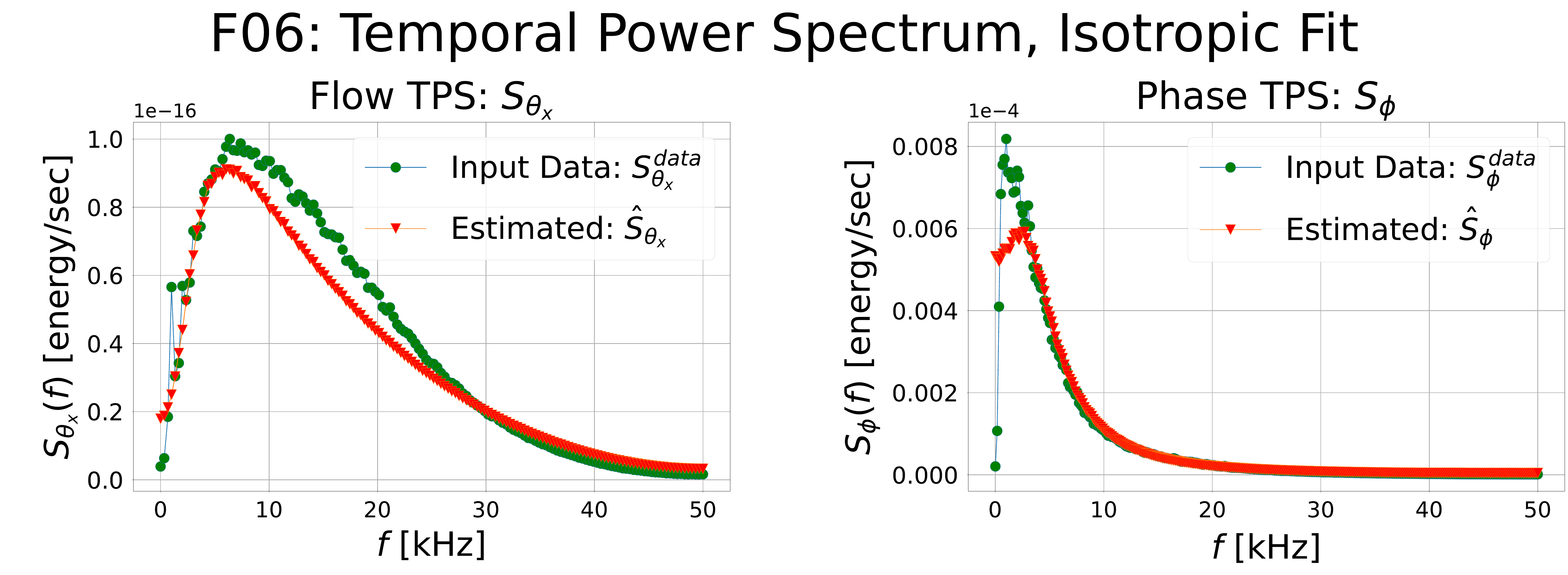}
    \includegraphics[width=0.6\textwidth]{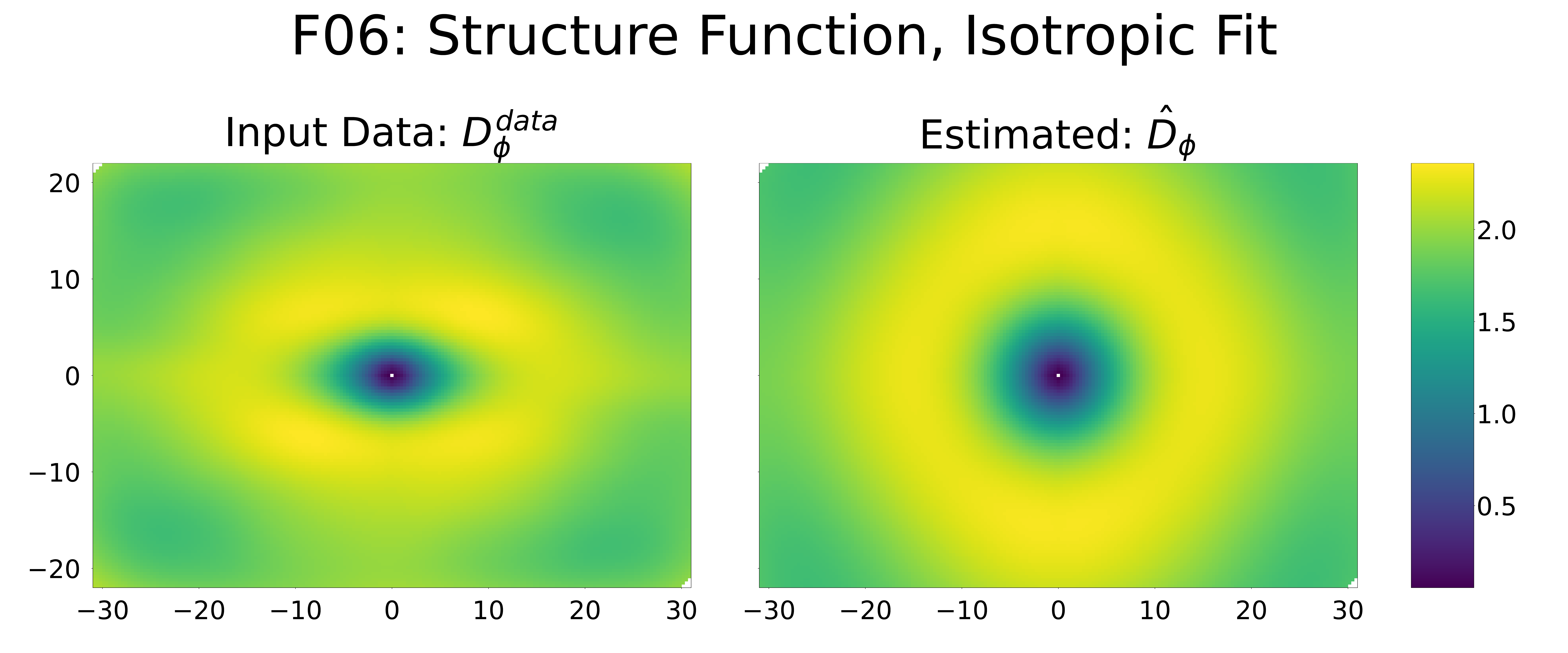}
    \includegraphics[width=0.7\textwidth]{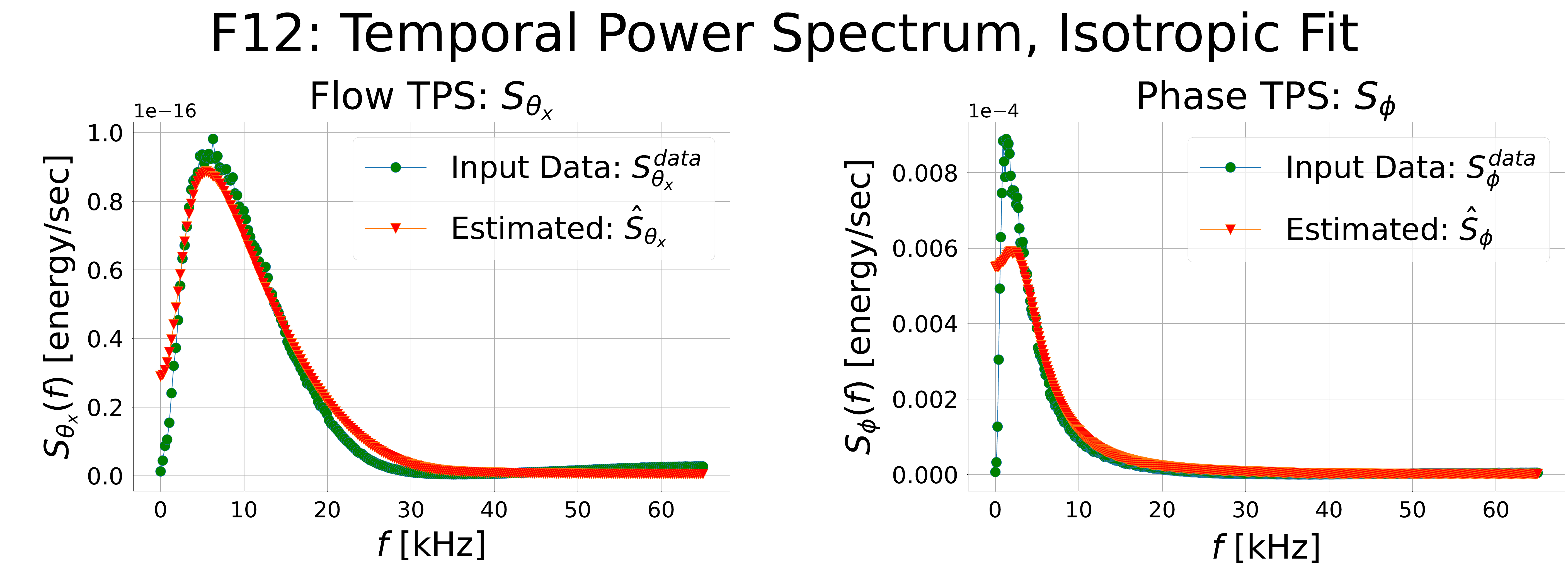}
    \caption{Isotropic estimation from measured data. {\bf Top:} Comparisons of power spectra obtained from measured data set F06 (blue) and from isotropic boiling flow data using estimated parameters (orange). The estimated flow TPS at top left matches the measured TPS reasonably well outside an underestimate in 5-25kHz, while the estimated phase TPS at top right matches at all but the lowest frequencies.
    {\bf Second row:} Comparison of the 2D structure functions~(\ref{eq: Estimate of Two-Dimensional Function}) obtained from measured data set F06 (left) and from  isotropic boiling flow data using estimated parameters (right). The estimated structure function gives an isotropic fit to the measured structure function but does not match its anisotropic statistics.
    {\bf Third row:} Results analogous to the top row but for data set F12.  Again the TPS fit is good except some mid-range frequencies for the flow TPS and low frequencies for the phase TPS. {\bf Not shown:} The structure function for F12 from estimated isotropic data shows a similar mismatch as for F06.}
    \label{fig: Isotropic Boiling Flow Results}
\end{figure}

Figure~\ref{fig: Isotropic Boiling Flow Results} shows the TPS and structure function results for measured data sets F06 and F12, again using isotropic boiling flow. Here, we plot the flow TPS as a function of frequency $f$ [kHz]; plots of the pre-multiplied flow TPS as a function of Strouhal number can be found in Appendix~\ref{appendix: TPS Plots with Strouhal Number}. In each case, the estimated flow TPS closely matches the flow TPS of the measured data. Further, the estimated phase TPS closely matches the measured phase TPS at frequencies above 10 kHz. However, the contours of the estimated structure function are circles, while the contours of the measured structure function are ellipses. 

Table~\ref{tab: Error Metrics} shows the error metrics for each data set under the isotropic boiling flow approximation. Our parameter estimation algorithm for isotropic phase screens matches the flow and phase TPS of the measured aero-optic data within 5\% to 12\% NRMSE but has roughly 30\% NRMSE for the 2D structure function. Thus, isotropic boiling flow captures the temporal statistics of the data reasonably well both visually and quantitatively, but does not capture the anisotropic spatial correlations of aero-optic phase screen data.

\begin{table}[htbp]
    \caption{Error Metrics from Measured Data - Isotropic Boiling Flow}
    \label{tab: Error Metrics}
    \centering
    \begin{tabular}{|c||c|c|}
        \hline
        \diagbox{NRMSE}{Data Set} & F06 & F12 \\
        \hline
        Flow TPS & 8.36\% & 5.40\% \\
        Phase TPS & 8.78\% & 11.67\% \\
        Structure Function & 32.64\% & 28.60\% \\
        \hline
    \end{tabular}
\end{table}

\subsubsection{Anisotropic estimation}
Table~\ref{tab: Anisotropic Parameter Estimates} shows the parameter estimates obtained using anisotropic boiling flow to fit each measured data set. Since the anisotropic and isotropic parameter estimation algorithms differ only in the estimation of $r_0$ and $\gamma_0$, the estimates of $L_0, \bm{v},$ and $\alpha$ are identical to those in Table~\ref{tab: Isotropic Parameter Estimates}. The low values of $\hat{\gamma}_0$ indicate that the spatial correlation length scale is much smaller in the $y$-axis than in the $x$-axis for both measured data sets. Further, since the algorithm for estimating $r_0$ depends on the value of $\hat{\gamma}_0$, the resulting $\hat{r}_0$ values for anisotropic boiling flow vary significantly from the estimates for isotropic boiling flow.

\begin{table}[htbp]
    \centering
    \caption{Parameter Estimates from Measured Data - Anisotropic Boiling Flow}
    \begin{tabular}{|c||c|c|c|c|c|}
        \hline
        \diagbox{Data Set}{Parameter} & $\hat{r}_0$ [mm] & $\hat{\gamma}_0$ & $\hat{L}_0$ & $\hat{\bm{v}}$ & $\hat{\alpha}$  \\
        \hline 
        F06 & 320.46 & 0.18 & \multicolumn{3}{c|}{\multirow{2}{*}{Same as Table~\ref{tab: Isotropic Parameter Estimates}}} \\
        F12 & 235.60 & 0.31 & \multicolumn{3}{c|}{} \\
        \hline 
    \end{tabular}
    \label{tab: Anisotropic Parameter Estimates}
\end{table}

\begin{figure}[h]
    \centering
    \includegraphics[width=0.7\textwidth]{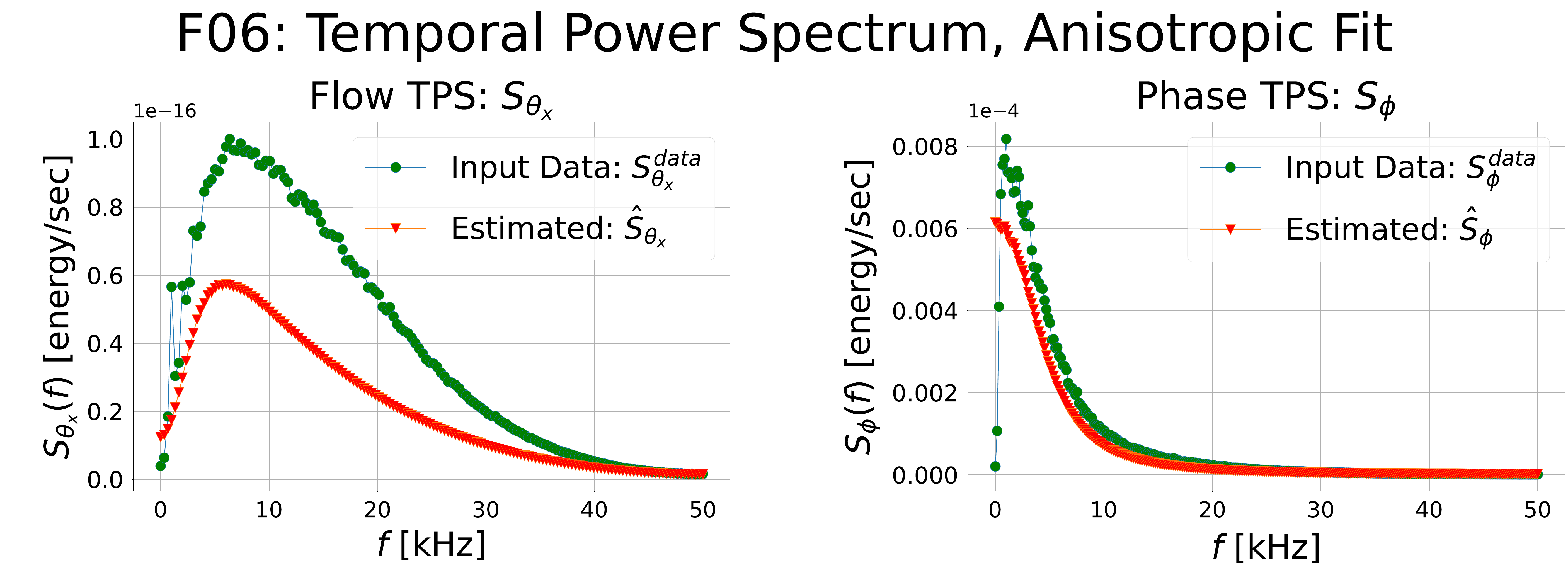}
    \includegraphics[width=0.6\textwidth]{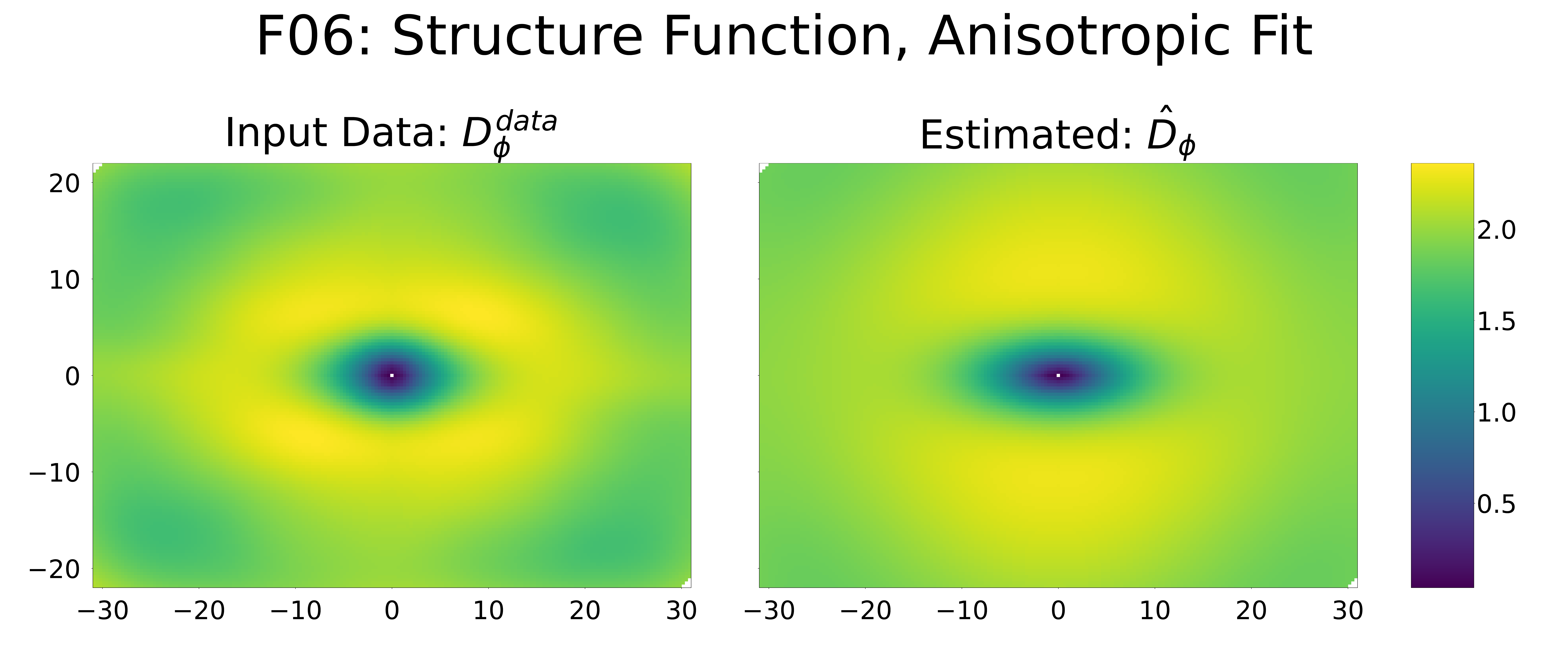}
    \caption{Anisotropic estimation from measured data. {\bf Top:} Comparisons of power spectra obtained from measured data set F06 (blue) and from anisotropic boiling flow data using estimated parameters (orange). The estimated TPS matches the reference TPS for both flow, $S_{\theta_x}$, and phase, $S_\phi$, at high frequencies but does not match either reference TPS at low and mid-range frequencies.
    {\bf Bottom:} Comparison of the structure functions~(\ref{eq: Estimate of Two-Dimensional Function}) obtained from measured data set F06 (left) and from anisotropic boiling flow data using estimated parameters (right). The estimated structure function matches the reference structure function, $D_{\phi}$ better than the isotropic estimates in Fig.~\ref{fig: Isotropic Boiling Flow Results}}
    \label{fig: Data Set F06 Anisotropic}
\end{figure}

\begin{figure}[h]
    \centering
    \includegraphics[width=0.7\textwidth]{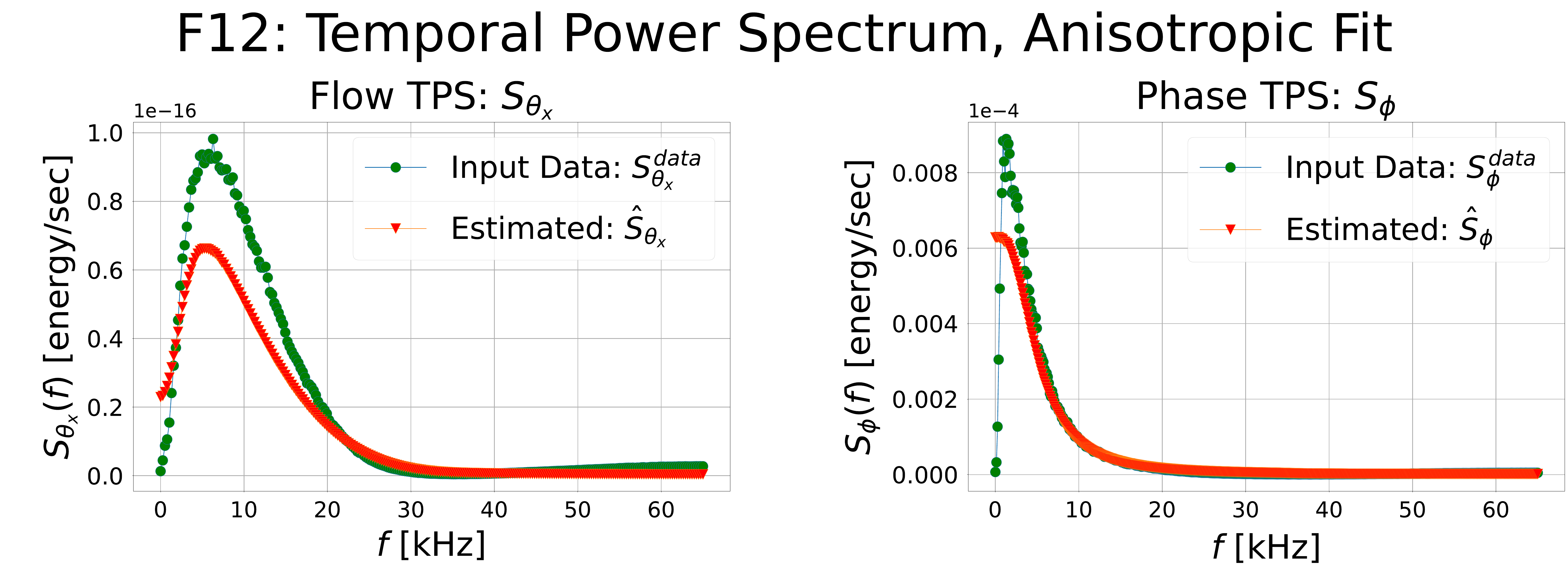}
    \includegraphics[width=0.6\textwidth]{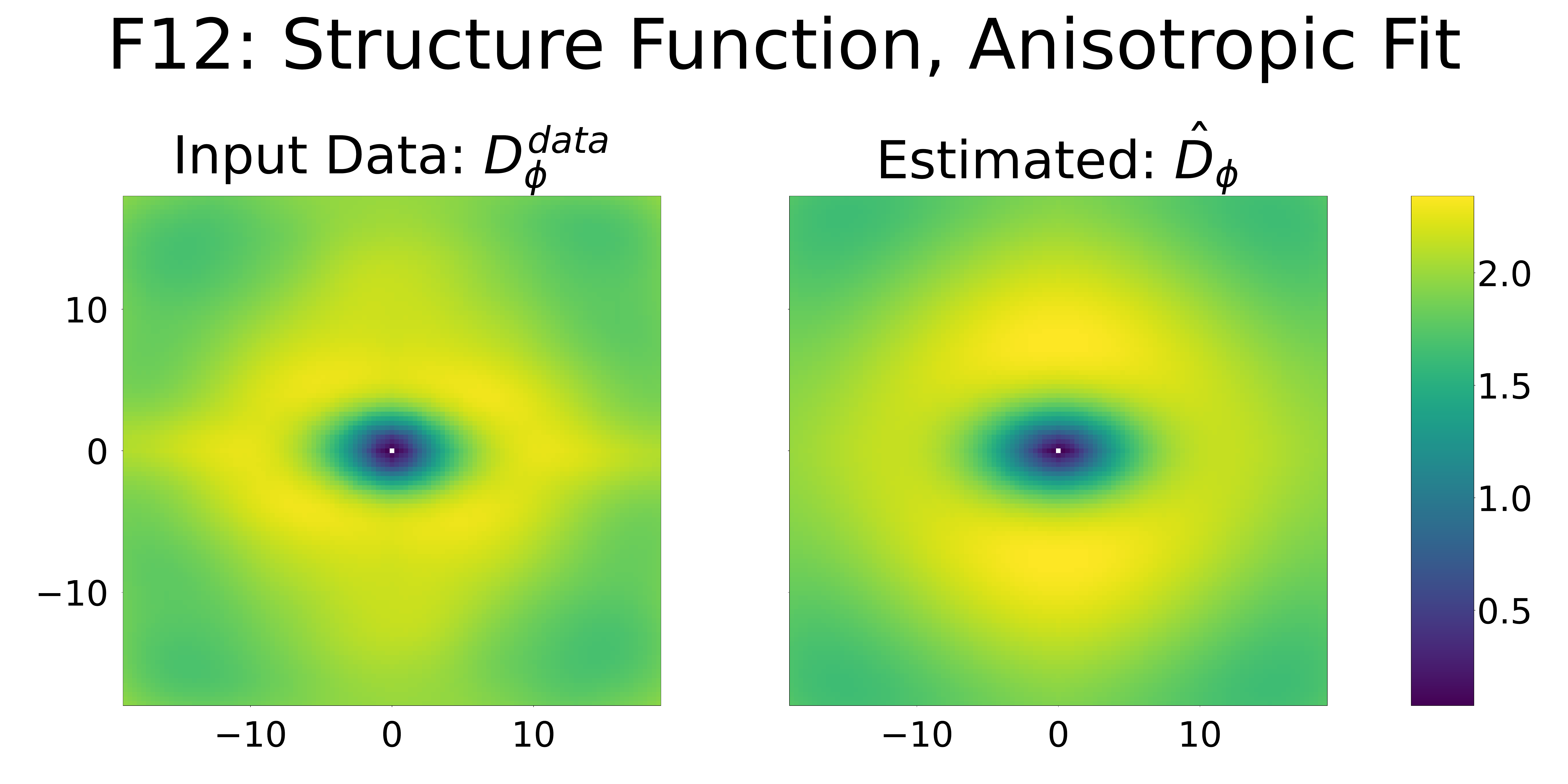}
    \caption{Results analogous to Fig.~\ref{fig: Data Set F06 Anisotropic} but for data set F12.}
    \label{fig: Data Set F12 Anisotropic}
\end{figure}

Figures~\ref{fig: Data Set F06 Anisotropic} and~\ref{fig: Data Set F12 Anisotropic} show the TPS and structure function results for measured data sets F06 and F12, respectively, again using anisotropic boiling flow. As with Fig.~\ref{fig: Isotropic Boiling Flow Results}, we plot the flow TPS as a function of frequency $f$ [kHz]; the pre-multiplied flow TPS as a function of Strouhal number is plotted in Appendix~\ref{appendix: TPS Plots with Strouhal Number}. In this case, neither the estimated flow TPS nor estimated phase TPS match the measured TPS as closely as isotropic boiling flow. However, the estimated structure functions more closely match the measured structure functions for both data sets, especially at low separation distances. Specifically, the elliptical contours of the estimated structure function more closely match those of the measured structure function. This suggests that anisotropic boiling flow captures the anisotropic spatial correlations of aero-optic phase screen data but do not match the TPS.

Table~\ref{tab: Anisotropic Error Metrics} shows the error metrics for each data set using anisotropic boiling flow. We first note that our parameter estimation algorithm for anisotropic phase screens has 11-27\% NRMSE for the flow and phase TPS, significantly larger than isotropic boiling flow. However, it has a structure function NRMSE between 18-26\%, much lower than isotropic boiling flow. Thus, while anisotropic boiling flow does not match the temporal statistics of the measured data, it matches the structure function more closely than isotropic boiling flow and better captures the anisotropic spatial correlations of the aero-optic data.

\begin{table}[htbp]
    \caption{Error Metrics from Measured Data - Anisotropic Boiling Flow}
    \label{tab: Anisotropic Error Metrics}
    \centering
    \begin{tabular}{|c||c|c|}
        \hline
        \diagbox{NRMSE}{Data Set} & F06 & F12 \\
        \hline
        Flow TPS & 26.10\% & 12.27\% \\
        Phase TPS & 11.28\% & 12.67\% \\
        Structure Function & 25.25\% & 18.23\% \\
        \hline
    \end{tabular}
\end{table}

\section{Conclusions}
\label{s: Conclusions}
In this paper, we introduced a method for applying a well-known atmospheric phase screen generation method, boiling flow, to aero-optic phase screen generation. We (1) introduced an algorithm for estimating the parameters of boiling flow from measured phase screen data and (2) generalized boiling flow to generate spatially anisotropic phase screens. This method can generate arbitrary-duration time series of spatially isotropic or anisotropic aero-optic phase screens using few parameters, offering a less expensive approach than physical experiments or computational fluid dynamics simulations.

We tested this method on both isotropic simulated boiling flow data with known parameter values and anisotropic measured aero-optic data \cite{Kemnetz}. The results indicate that our method matches both the temporal power spectrum (TPS) and 2D phase structure function of the isotropic simulated data. Furthermore, while this model does not closely match both the TPS and structure function of the measured data simultaneously, it can match reasonably well either the TPS using isotropic boiling flow or the structure function using anisotropic boiling flow.

\appendix 

\section{Bound for $T_{\max}$ in the Flow Velocity Estimation Algorithm}\label{appendix: Tolerance}
In this section, we derive the lower bound for the peak cross-correlation estimate $\hat{R}_{\phi'}(\hat{\bm{v}};T)$ enforced for each time-lag $T$ in the flow velocity estimation algorithm in Sec.~\ref{s: Estimation of Temporal Parameters}:
\begin{align}\label{eq: T Lower Bound}
    \hat{R}_{\phi'}(\hat{\bm{v}}; T) \geq \frac{15\sqrt{2}}{\sqrt{N_T-T}}.
\end{align}
We show that, if this bound holds, then the SNR of the estimate $\hat{R}_{\phi'}(\hat{\bm{v}};T)$ is at least $10$, with $5\sigma$ confidence. Requiring that this SNR is at least 10 significantly reduces the probability that the flow velocity estimate $\hat{\bv}(T)$ is obscured by boiling. Enforcing this bound for each time-lag $T$ is equivalent to setting
\begin{align}
    T_{\max} \leq \max\Bigl\{T: \hat{R}_{\phi'}(\hat{\bm{v}}; T) \geq \frac{15\sqrt{2}}{\sqrt{N_T-T}}\Bigr\}
\end{align}
in the flow velocity estimation algorithm.

Note that, for $n \geq T$, Eq.~(\ref{eq: Boiling Flow}) has an image space representation:
\begin{align}\label{eq: Image Space Boiling Flow}
    \phi'_n(\bm{i}) = \alpha^T\phi'_{n-T}(\bm{i}-T\bm{v}^{\text{gt}})+\sqrt{1-\alpha^2}\sum_{t=0}^{T-1}\alpha^t B'_{n-t}(\bm{i}-t \hspace{0.05cm}\bm{v}^{\text{gt}}).
\end{align}
Here, $\phi_n'$ denotes the the mean-subtracted $\phi_n$ divided by its standard deviation, $B_n'$ denotes the corresponding (normalized) image-space boiling component~(\ref{eq: Boiling}), and $\bm{v}^{\text{gt}}$ is the ground-truth flow velocity. Note that, while the components of $\bm{i}-t \bm{v}^{\text{gt}}$ may not be integers, these positions are still well-defined given the spatial grid spacing $\Delta$. Given the assumptions of boiling flow and the effect of TTP removal, it follows that both $\phi_n'$ and $B_n'$ follow a zero-mean, wide-sense stationary multivariate Gaussian distribution (assuming the effect of tip/tilt removal on the distribution of each object is minimal), where each component has unit variance. Furthermore, $B_m'$ is independent of $\phi'_n$ for all $m > n$.

We compute the SNR for an input $\bm{v}$ and time-lag $T$, which we denote SNR$(\bm{v}, T)$, as the ratio between the mean and standard deviation of the estimate $\hat{R}_{\phi'}(\bm{v}; T)$ of Eq.~(\ref{eq: Cross-Correlation}). We compute this estimate using an average over the time series and over the pixel pairs in the aperture, so the mean of this estimate $\mu_{\hat{R}}(\bm{v}; T)$ is equal to $R_{\phi'}(\bm{v}; T)$. We use Eq.~(\ref{eq: Image Space Boiling Flow}), the assumed statistical properties of $\phi_n'$ and $B_n'$, and an additional assumption that $\phi'_{n-T}(\bm{i}-T\bm{v})\phi'_n(\bm{i})$ and $\phi'_{m-T}(\bm{i}-T\bm{v})\phi'_m(\bm{i})$ are independent for $n \neq m$ to derive an upper bound for the variance of this estimate:
\begin{align}
    \Var[\hat{R}_{\phi'}(\bm{v}; T)] &\leq \frac{1}{N_T-T}\Var[\phi'_n(\bm{i})\phi'_{n-T}(\bm{i}-\bm{v})] \\ \label{eq: Variance Form}
    &= \frac{1}{N_T-T}\left(1+R^2_{\phi'}(\bm{v}; T)\right).
\end{align}
Here, $N_T$ is the number of time-steps of training data. Importantly, this upper bound applies for all inputs $\bm{v}$.

Defining $\sigma_{\hat{R}}(\bm{v}; T)=\sqrt{\Var[\hat{R}_{\phi'}(\bm{v}; T)]}$, we may then derive a lower bound for SNR$(\bm{v}, T)$:
\begin{align}
    \text{SNR}(\bm{v}, T) &= \frac{\mu_{\hat{R}}(\bm{v}; T)}{\sigma_{\hat{R}}(\bm{v}; T)} \\
    &\geq \sqrt{N_T-T} \times \frac{R_{\phi'}(\bm{v}; T)}{\sqrt{1+R^2_{\phi'}(\bm{v}; T)}}.
\end{align}
Since $R_{\phi'}(\bm{v};T)\leq 1$ for all $\bm{v}$,
\begin{gather} \label{eq: Std Dev Bound}
    \sigma_{\hat{R}}(\bm{v}; T) \leq \frac{\sqrt{2}}{\sqrt{N_T-T}}, \\
    \text{SNR}(\bm{v}, T) \geq \sqrt{N_T-T}\times \frac{R_{\phi'}(\bm{v}; T)}{\sqrt{2}}.
\end{gather}
Thus, if
\begin{align}\label{eq: Correlation Bound}
    R_{\phi'}(\bm{v}; T) \geq \frac{10\sqrt{2}}{\sqrt{N_T-T}},
\end{align}
then the SNR of the estimate $\hat{R}_{\phi'}(\bm{v}; T)$ is at least 10.

We wish to enforce $\text{SNR}(\hat{\bm{v}}, T)\geq 10$ for each $T$. That is, we require only that the peak cross-correlation estimate $\hat{R}_{\phi'}(\hat{\bm{v}};T)$ has an SNR of at least 10. To accomplish this, we enforce the bound~(\ref{eq: Correlation Bound}) for $\bm{v}=\hat{\bm{v}}$ at each $T$. Since we do not have access to the value of $R_{\phi'}(\bm{v}; T)$,  we substitute our estimate $\hat{R}_{\phi'}(\hat{\bm{v}}; T)$ into the LHS of~(\ref{eq: Correlation Bound}) with $5\sigma$ confidence:
\begin{align}
    \hat{R}_{\phi'}(\hat{\bm{v}}; T)\pm 5\sigma_{\hat{R}}(\hat{\bm{v}}; T)\geq \frac{10\sqrt{2}}{\sqrt{N_T-T}}.
\end{align}
Given the bound on the standard deviation~(\ref{eq: Std Dev Bound}), we may conclude that if (\ref{eq: T Lower Bound}) holds,
then~(\ref{eq: Correlation Bound}) holds for $\bm{v}=\hat{\bm{v}}$ with $5\sigma$ confidence.

\section{Pre-Processing Measured Data Sets}\label{appendix: Pre-Processing}
We applied FIR filters to the measured data sets F06 and F12 before applying our parameter estimation algorithm. We used both notch filters and band-stop filters to remove low-frequency peaks in the TPS of each data set. 

The notch filters use a modulated hamming window to reduce the TPS value at a given frequency $f_0$ [Hz] by a factor $r\in(0,1)$. To construct a notch filter, we first generate a Hamming window $h_n$ of length $N_W$ and compute
\begin{align}
	g_n = h_n \ast \cos\left(2\pi \frac{f_0}{f_s}n\right),
\end{align}
where $f_s$ [Hz] is the temporal sampling frequency of the data set (listed in Table~\ref{tab: Experimental Data Sets}). We then normalize $g_n$ by its frequency response at $f_0$:
\begin{align}
    \tilde{g}_n = \frac{g_n}{\sum_{m=0}^{N_W-1}g_m\cos\left(2\pi \frac{f_0}{f_s}m\right)}.
\end{align}
Finally, given a ratio $r \in (0,1)$, we use the impulse response
\begin{align}
	\delta_n - (1-\sqrt{r})\tilde{g}_n
\end{align}
for the notch FIR filter.

Similarly, the band-stop filters use a modulated sinc function to reduce the TPS values in a frequency range $f \in (f_1, f_2)$ [Hz] by a factor of $r\in(0,1)$. To construct this filter, we first identify the center frequency $f_0 = \frac{1}{2}(f_1+f_2)$ and frequency range $f_r = \frac{1}{2}(f_2-f_1)$ and then generate a scaled sinc function
\begin{align}
    g_n = 2f_r\; \text{sinc}(2\pi f_r n)
\end{align}
of length $N_W$. Next, we apply a Hamming window $h_n$ of the same length to obtain $\tilde{g}_n = h_n g_n$ and modulate the windowed sinc function via
\begin{align}
    b_n = \tilde{g}_n \ast \cos\left(2\pi \frac{f_0}{f_s}n\right).
\end{align}
As with the notch filter, we then normalize the modulated sinc function via
\begin{align}
    \tilde{b}_n = \frac{b_n}{\sum_{m=0}^{N_W-1}b_m\cos\left(2\pi \frac{f_0}{f_s}m\right)}.
\end{align}
Finally, we compute
\begin{align}
    \delta_n - (1-\sqrt{r})\tilde{b}_n,
\end{align}
the impulse response of the band-stop filter.

Tables~\ref{tab: F06 Pre-Processing Parameters} and~\ref{tab: F12 Pre-Processing Parameters} list the parameters of each filter applied to F06 and F12, respectively. The filter length $N_W$ determines the bandwidth of the notch filter and the frequency resolution of both filters.

\begin{table}[htbp]
    \centering
    \caption{Parameters of F06 Pre-Processing Filters}
    \label{tab: F06 Pre-Processing Parameters}
    \begin{tabular}{|c|c|c|c|c|}
        \hline
        Filter Type & $N_W$ & $f_r$ & $f_0$ & $r$ \\
        \hline
        Band-Stop & 2001 & 350 & 850 & 0.81 \\
        Notch & 1029 & N/A & 683.59 & 0.81 \\
        Notch & 2049 & N/A & 976.56 & 0.81 \\
        \hline 
    \end{tabular}
\end{table}

\begin{table}[htbp]
    \centering
    \caption{Parameters of F12 Pre-Processing Filters}
    \label{tab: F12 Pre-Processing Parameters}
    \begin{tabular}{|c|c|c|c|c|}
        \hline
        Filter Type & $N_W$ & $f_r$ & $f_0$ & $r$ \\
        \hline
        Band-Stop & 2001 & 300 & 1000 & 0.81\\
        Notch & 2049 & N/A & 900 & 0.95 \\
        \hline
    \end{tabular}
\end{table}

\section{Pre-Multiplied Flow TPS Plots as a Function of Strouhal Number}\label{appendix: TPS Plots with Strouhal Number}
In this section, we show the pre-multiplied flow TPS as a function of the Strouhal number $\text{St}_{\delta}$ for both measured data sets. The Strouhal number is a unit-less quantity given by
\begin{align}
    \text{St}_{\delta} = \frac{f \times \delta^*}{U_c},
\end{align}
where $f$ is the temporal frequency in Hz, $\delta^*$ is the boundary layer thickness in meters, and $U_c$ is the convective velocity in meters per second. For both measured data sets F06 and F12, we used $\delta^* = 15.6$ mm, as estimated in Refs.~\citenum{Kemnetz, KemnetzDissertation}. We computed $U_c=\hat{v}_x \Delta f_s$ using the flow velocity estimates $\hat{v}_x$ from Table~\ref{tab: Isotropic Parameter Estimates} and pixel spacing $\Delta$ and sampling frequency $f_s$ from Table~\ref{tab: Experimental Data Sets}; this resulted in values $U_c=172.37$ for F06 and $U_c=160.26$ for F12.

\begin{figure}
    \centering
    \includegraphics[width=0.8\textwidth]{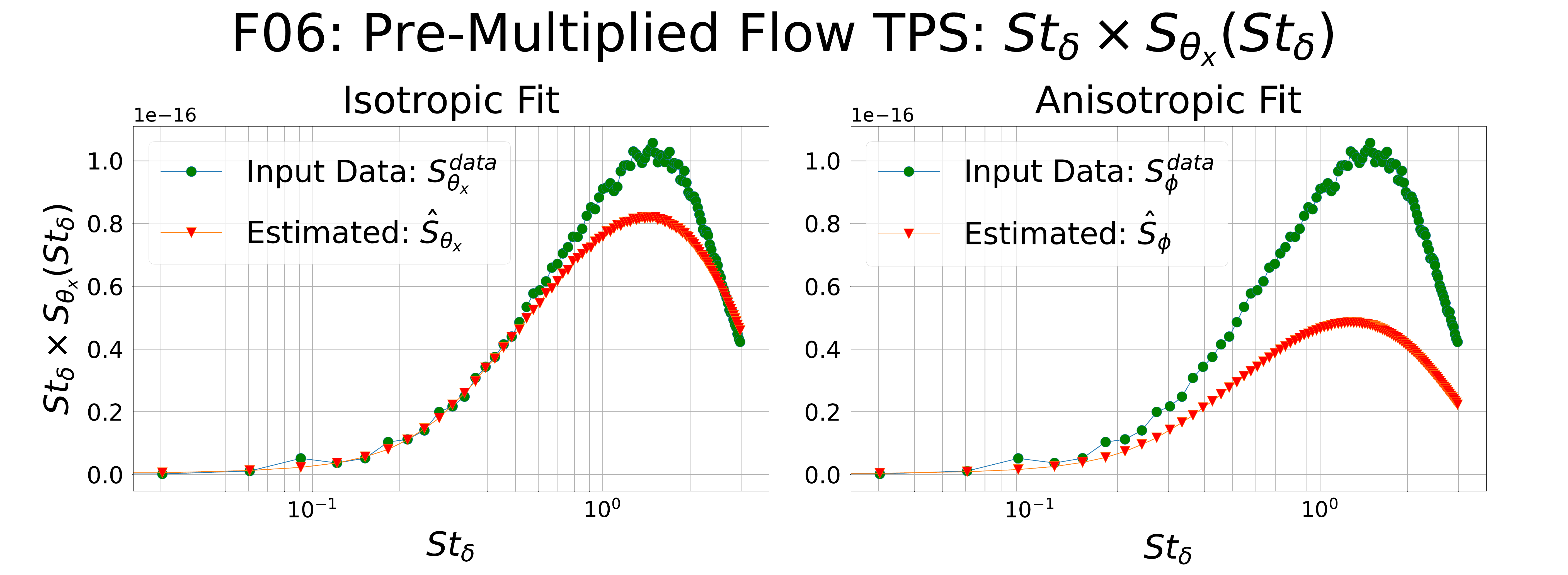}
    \includegraphics[width=0.8\textwidth]{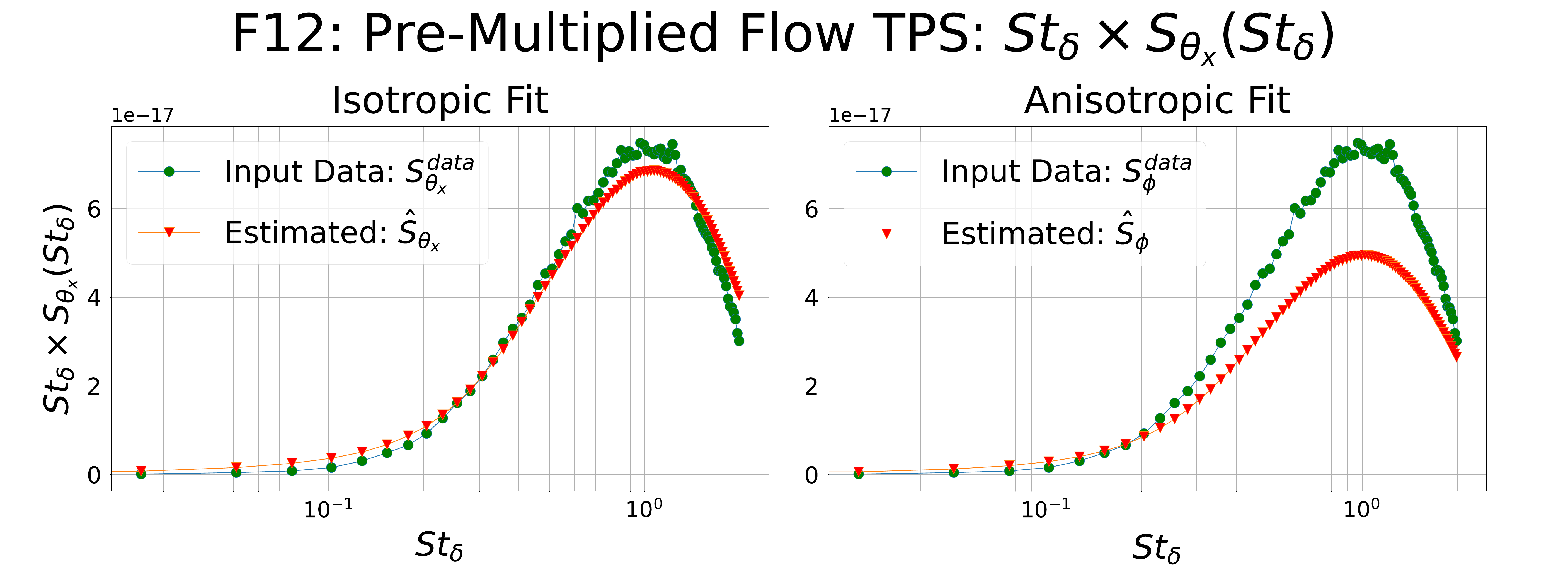}
    \caption{Comparisons of the pre-multiplied flow temporal power spectrum (TPS) as a function of Strouhal number $\text{St}_{\delta}$ obtained from measured data (blue) and from boiling flow data using estimated parameters (orange). {\bf Top:} data set F06. {\bf Bottom:} data set F12. Results include both isotropic boiling flow (left) and anisotropic boiling flow (right). }
    \label{fig: Pre-Multipled Flow TPS}
\end{figure}

Figure~\ref{fig: Pre-Multipled Flow TPS} shows the pre-multiplied flow TPS results as a function of the Strouhal number $\text{St}_{\delta}$ for both measured data sets F06 and F12. We include results for both isotropic and anisotropic boiling flow. The $y$-axis of each plot shows the flow TPS multiplied by the Strouhal number input, $\text{St}_{\delta}\times S_{\theta_x}(\text{St}_{\delta})$. Further, in these plots, we include pre-multiplied flow TPS values up to $\text{St}_\delta=3$ for F06 and up to $\text{St}_\delta=2$ for F12; we exclude the highest frequencies since the measured data sets appeared to be aliased. These results illustrate a similar fit to the flow TPS as the results in Figs.~\ref{fig: Isotropic Boiling Flow Results}, ~\ref{fig: Data Set F06 Anisotropic}, and ~\ref{fig: Data Set F12 Anisotropic}.

\section*{Disclosures}
The views expressed are those of the author and do not necessarily reflect the official policy or position of the Department of the Air Force, the Department of Defense, or the U.S. Government.  Approved for public release; distribution is unlimited.  Public Affairs release approval \# \PAnumber.

\section* {Code, Data, and Materials Availability} 
The code used to generate the results and figures is available in a Github repository linked in Ref.~\citenum{Repo}.

The measured data sets used in this article were taken at the University of Notre Dame in the trisonic wind tunnel facility within the hessert laboratory for aerospace research. For more details concerning the experiment, please refer to Ref.~\citenum{Kemnetz}. This data is publicly available at the following link: https://www.datadepot.rcac.purdue.edu/bouman/.

\section* {Acknowledgments}
C.A.B. was partially supported by the Showalter Trust.  J.W.U., G.T.B., and C.A.B. were partially supported by AFRL/RDKL FA9451-20-2-0008. The authors would like to thank the Showalter family and the United States Air Force for supporting this research.

\bibliography{report}

@article{Srinath,
author = {Srikar Srinath and Lisa A. Poyneer and Alexander R. Rudy and S. Mark Ammons},
journal = {Opt. Express},
number = {26},
pages = {33335--33349},
publisher = {Optica Publishing Group},
title = {Computationally efficient autoregressive method for generating phase screens with frozen flow and turbulence in optical simulations},
volume = {23},
month = {Dec},
year = {2015},
url = {https://opg.optica.org/oe/abstract.cfm?URI=oe-23-26-33335},
note = {[doi:10.1364/OE.23.033335]},
}

@article{WangAero-Optics,
author = {Wang, Kan and Wang, Meng},
address = {Cambridge, UK},
copyright = {Copyright © Cambridge University Press  2012},
issn = {0022-1120},
journal = {Journal of Fluid Mechanics},
language = {eng},
pages = {122-151},
publisher = {Cambridge University Press},
title = {Aero-optics of subsonic turbulent boundary layers},
volume = {696},
year = {2012},
note = {[doi:10.1017/jfm.2012.11]}
}

@inproceedings{WangComputation,
author = {Kan Wang and Meng Wang and Stanislav Gordeyev and Eric Jumper},
booktitle = {41st Plasmadynamics and Lasers Conference},
language = {eng},
title = {Computation of Aero-Optical Distortions over a Cylindrical Turret with Passive Flow Control},
year = {2010},
pages={4498},
publisher = {American Institute of Aeronautics and Astronautics},
note = {[doi:10.2514/6.2010-4498]}
}

@article{Porter,
author = {Porter, Chris and Rennie, Mark and Jumper, Eric},
address = {Virginia},
copyright = {Copyright © 2014 by Chris O. Porter, R. Mark Rennie, and Eric J. Jumper. Published by the American Institute of Aeronautics and Astronautics, Inc., with permission. Copies of this paper may be made for personal or internal use, on condition that the copier pay the $10.00 per-copy fee to the Copyright Clearance Center, Inc., 222 Rosewood Drive, Danvers, MA 01923; include the code and $10.00 in correspondence with the CCC.},
issn = {0001-1452},
journal = {AIAA journal},
language = {eng},
number = {3},
pages = {532-541},
publisher = {American Institute of Aeronautics and Astronautics},
title = {Computation of the Aero-Optical Environment of a Helicopter Using Prescribed-Wake Methods},
volume = {53},
year = {2015},
note = {[doi:10.2514/1.J052969]}
}

@inproceedings{Kemnetz,
	author={Matthew R. Kemnetz and Stanislav Gordeyev},
	title={Optical investigation of large-scale boundary-layer structures},
	year={2016},
    publisher = {American Institute of Aeronautics and Astronautics},
	eventdate={4 - 8 Jan 2016},
    pages = {1460},
	booktitle= {54th AIAA Aerospace Sciences Meeting},
	location={San Diego, California},
    note = {[doi:10.2514/6.2016-1460]}
}

@inbook{Schmidt,
	author = {Jason D. Schmidt},
	title = {Numerical Simulation of Optical Wave Propagation with Examples in MATLAB},
	copyright = {Copyright © 2010 Society of Photo-Optical Instrumentation Engineers All rights reserved. No part of this publication may be reproduced or distributed in any form or by any means without written permission of the publisher.},
	address = {United States},
	publisher = {SPIE},
    volume = {PM199},
    isbn = {0819483265},
    language = {eng},
    pages = {149-184},
    chapter = {9},
	year = {2010},
	month = {July},
	date = {2010-7-29},
	note = {[doi:10.1117/3.866274.ch9]},
	url = {https://doi.org/10.1117/3.866274}
}

@ARTICLE{Welch,

  author={Welch, P.},

  journal={IEEE Transactions on Audio and Electroacoustics}, 

  title={The use of fast Fourier transform for the estimation of power spectra: A method based on time averaging over short, modified periodograms}, 

  year={1967},

  volume={15},

  number={2},

  pages={70-73},

  note={[doi:10.1109/TAU.1967.1161901]},
}

@article{Vogel,
    author = {Curtis R. Vogel and Glenn A. Tyler and Donald J. Wittich},
    journal = {J. Opt. Soc. Am. A},
    number = {7},
    pages = {1666--1679},
    publisher = {Optica Publishing Group},
    title = {Spatial-temporal-covariance-based modeling, analysis, and simulation of aero-optics wavefront aberrations},
    volume = {31},
    month = {Jul},
    year = {2014},
    url = {https://opg.optica.org/josaa/abstract.cfm?URI=josaa-31-7-1666},
    note = {[doi:10.1364/JOSAA.31.001666]},
}

@article{JumperPhysicsMeasurement,
   author = "Jumper, Eric J. and Gordeyev, Stanislav",
   title = "Physics and Measurement of Aero-Optical Effects: Past and Present", 
   journal= "Annual Review of Fluid Mechanics",
   year = "2017",
   volume = "49",
   number={1},
   pages = "419-441",
   note = "[doi:10.1146/annurev-fluid-010816-060315]",
   url = "https://www.annualreviews.org/content/journals/10.1146/annurev-fluid-010816-060315",
   publisher = "Annual Reviews",
   issn = "1545-4479",
   type = "Journal Article",
}

@article{WangPhysicsComputation,
   author = "Wang, Meng and Mani, Ali and Gordeyev, Stanislav",
   title = "Physics and Computation of Aero-Optics", 
   journal= "Annual Review of Fluid Mechanics",
   year = "2012",
  volume={44},
  number={1},
   pages = "299-321",
   note = "[doi:10.1146/annurev-fluid-120710-101152]",
   url = "https://www.annualreviews.org/content/journals/10.1146/annurev-fluid-120710-101152",
   publisher = "Annual Reviews",
   issn = "1545-4479",
   type = "Journal Article",
}

@inproceedings{Siegenthaler,
author = {Siegenthaler, John P. and Jumper, Eric J. and Gordeyev, Stanislav},
booktitle = {46th AIAA Aerospace Sciences Meeting and Exhibit},
copyright = {Copyright 2020 Elsevier B.V., All rights reserved.},
isbn = {9781563479373},
language = {eng},
title = {Atmospheric Propagation vs. Aero-Optics},
pages={1076},
publisher = {American Institute of Aeronautics and Astronautics},
year = {2008},
note = {[doi:10.2514/6.2008-1076]}
}

@article{Holmes,
author = {Holmes, Richard B.},
address = {Virginia},
copyright = {Copyright © 2022 by the American Institute of Aeronautics and Astronautics, Inc. All rights reserved. All requests for copying and permission to reprint should be submitted to CCC at ; employ the eISSN to initiate your request. See also AIAA Rights and Permissions .},
issn = {0001-1452},
journal = {AIAA journal},
language = {eng},
number = {10},
pages = {5633-5644},
publisher = {American Institute of Aeronautics and Astronautics},
title = {Adaptive Optics for Directed Energy: Fundamentals and Methodology},
volume = {60},
year = {2022},
note = {[doi:10.2514/1.J061766]}
}

@inproceedings{SiegenthalerShear,
author = {Siegenthaler, John P. and Gordeyev, Stanislav and Jumper, Eric},
booktitle = {36th AIAA Plasmadynamics and Lasers Conference},
copyright = {Copyright 2013 Elsevier B.V., All rights reserved.},
isbn = {9781624100604},
language = {eng},
title = {Shear layers and aperture effects for aero-optics},
pages={4772},
year = {2005},
publisher = {American Institute of Aeronautics and Astronautics},
note = {[doi:10.2514/6.2005-4772]}
}

@inproceedings{Cress,
author = {Jacob Cress and Stanislav Gordeyev and Martiqua Post and Eric Jumper},
booktitle = {39th Plasmadynamics and Lasers Conference},
language = {eng},
title = {Aero-Optical Measurements in a Turbulent, Subsonic Boundary Layer at Different Elevation Angles},
publisher = {American Institute of Aeronautics and Astronautics},
pages={4214},
year={2005},
note = {[doi:10.2514/6.2008-4214]}
}

@article{Sahba,
author = {Shervin Sahba and Diya Sashidhar and Christopher C. Wilcox and Austin McDaniel and Steven L. Brunton and J. Nathan Kutz},
title = {{Dynamic mode decomposition for aero-optic wavefront characterization}},
volume = {61},
journal = {Optical Engineering},
number = {1},
publisher = {SPIE},
pages = {013105},
year = {2022},
note = {[doi:10.1117/1.OE.61.1.013105]},
URL = {https://doi.org/10.1117/1.OE.61.1.013105}
}

@inproceedings{Kutz,
author = {J. Nathan Kutz and Diya Sashidhar and Shervin Sahba and Steven L. Brunton and Austin McDaniel and Christopher C. Wilcox},
title = {{Physics-informed machine-learning for modeling aero-optics}},
volume = {11817},
booktitle = {Applied Optical Metrology IV},
editor = {Erik Novak and James D. Trolinger and Christopher C. Wilcox},
pages = {118170E},
year = {2021},
series = "Proc. SPIE", 
note = {[doi:10.1117/12.2596540]},
URL = {https://doi.org/10.1117/12.2596540}
}

@inproceedings{ShafferNeuralNetwork,
author = {Benjamin D. Shaffer and Jeremy R. Vorenberg and Christopher C. Wilcox and Austin J. McDaniel},
title = {{Neural network forecasting of transonic turbulent flow for adaptive optics control}},
volume = {12239},
booktitle = {Unconventional Imaging and Adaptive Optics 2022},
editor = {Jean J. Dolne and Mark F. Spencer},
series = "Proc. SPIE", 
pages = {122390H},
year = {2022},
note = {[doi:10.1117/12.2631995]},
URL = {https://doi.org/10.1117/12.2631995}
}

@inproceedings{BurnsALatency,
author = {Burns, Robert and Jumper, Eric and Gordeyev, Stanislav},
booktitle = {53rd AIAA Aerospace Sciences Meeting},
copyright = {Copyright 2016 Elsevier B.V., All rights reserved.},
isbn = {9781624103438},
language = {eng},
title = {A Latency-Tolerant Architecture for Airborne Adaptive Optic Systems},
pages={0679},
publisher = {American Institute of Aeronautics and Astronautics},
year = {2015},
note = {[doi:10.2514/6.2015-0679]}
}

@inproceedings{BurnsARobust,
author = {Burns, Robert and Jumper, Eric and Gordeyev, Stanislav},
booktitle = {47th AIAA Plasmadynamics and Lasers Conference},
copyright = {Copyright 2017 Elsevier B.V., All rights reserved.},
isbn = {9781624104343},
language = {eng},
title = {A Robust Modification of a Predictive Adaptive-Optic Control Method for Aero-Optics},
pages={3529},
year = {2016},
publisher = {American Institute of Aeronautics and Astronautics},
note = {[doi:10.2514/6.2016-3529]}
}

@article{Martin,
  title={GSM: a Grating Scale Monitor for atmospheric turbulence measurements. I. The instrument and first results of angle of arrival measurements.},
  author={Martin, F and Tokovinin, A and Agabi, A and Borgnino, J and Ziad, AGSM},
  journal={Astronomy and Astrophysics Suppl. Ser.},
  volume={108},
  pages={173--180},
  year={1994}
}

@inproceedings{GordeyevFluidDynamics,
author = {Gordeyev, Stanislav and Jumper, Eric},
booktitle = {40th AIAA Plasmadynamics and Lasers Conference},
copyright = {Copyright 2020 Elsevier B.V., All rights reserved.},
isbn = {1563479753},
language = {eng},
title = {Fluid Dynamics and Aero-Optical Environment Around Turrets},
publisher = {American Institute of Aeronautics and Astronautics},
year = {2009},
pages={4224},
note = {[doi:10.2514/6.2009-4224]}
}

@article{GordeyevFluidDynamics2,
author = {Gordeyev, Stanislav and Jumper, Eric},
copyright = {2010 Elsevier Ltd},
issn = {0376-0421},
journal = {Progress in Aerospace Sciences},
language = {eng},
number = {8},
pages = {388-400},
publisher = {Elsevier Ltd},
title = {Fluid dynamics and aero-optics of turrets},
volume = {46},
year = {2010},
note = {[doi:10.1016/j.paerosci.2010.06.001]}
}

@inproceedings{Utley,
author = {Jeffrey W. Utley and Gregery T. Buzzard and Charles A. Bouman and Matthew R. Kemnetz},
title = {{Data-driven synthetic wavefront generation for boundary layer data}},
volume = {13149},
booktitle = {Unconventional Imaging, Sensing, and Adaptive Optics 2024},
editor = {Jean J. Dolne and Santasri R. Bose-Pillai and Matthew Kalensky},
series = {Proc. SPIE},
pages = {131490A},
year = {2024},
note = {[doi:10.1117/12.3027740]},
URL = {https://doi.org/10.1117/12.3027740}
}

@article{Shaffer,
author = {Shaffer, Benjamin D. and McDaniel, Austin J. and Wilcox, Christopher C. and Ahn, Edwin S.},
address = {Washington},
copyright = {Copyright 2021 Elsevier B.V., All rights reserved.},
issn = {1559-128X},
journal = {Appl. Opt.},
language = {eng},
number = {25},
pages = {G170--G180},
publisher = {Optica Publishing Group},
title = {Dynamic mode decomposition based predictive model performance on supersonic and transonic aero-optical wavefront measurements},
volume = {60},
month = {Sep},
year = {2021},
note = {[doi:10.1364/AO.426031]}
}

@inproceedings{ShafferPredictive,
author = {Benjamin D. Shaffer and Austin J. McDaniel and Christopher C. Wilcox},
title = {{Predictive modeling of wavefront error using dynamic mode decomposition}},
volume = {11490},
booktitle = {Interferometry XX},
editor = {Michael B. North Morris and Katherine Creath and Rosario Porras-Aguilar},
series = {Proc. SPIE},
pages = {114900E},
year = {2020},
note = {[doi:10.1117/12.2569869]},
URL = {https://doi.org/10.1117/12.2569869}
}

@inproceedings{Visbal,
author = {Visbal, Miguel R. and Rizzetta, Donald P.},
booktitle = {46th AIAA Aerospace Sciences Meeting and Exhibit},
copyright = {Copyright 2020 Elsevier B.V., All rights reserved.},
isbn = {9781563479373},
language = {eng},
title = {Effect of Flow Excitation on Aero-Optical Aberration},
location = {Reno, Nevada},
year = {2008},
pages={1074},
publisher = {American Institute of Aeronautics and Astronautics},
note = {[doi:10.2514/6.2008-1074]}
}

@inbook{Tatarski,
author = {Tatarski, V. I.},
year = {1961},
chapter = {4},
pages = {59--80},
publisher = {Dover},
address = {Mineola, New York},
title = {Wave Propagation in a Turbulent Medium},
languate={Russian, English},
translator = {Richard A. Silverman},
note = {English translation of the original Russian edition},
}

@article{Fitzgerald,
author = {Fitzgerald, E. J. and Jumper, E. J.},
address = {Cambridge, UK},
copyright = {2004 Cambridge University Press},
issn = {0022-1120},
journal = {Journal of fluid mechanics},
language = {eng},
pages = {153-189},
publisher = {Cambridge University Press},
title = {The optical distortion mechanism in a nearly incompressible free shear layer},
volume = {512},
year = {2004},
note = {[doi:10.1017/S0022112004009553]}
}

@inbook{Chernov,
    title = {Wave Propagation in a Random Medium},
    author = {Chernov, Lev A.},
    translator = {Silverman, Richard A.},
    year = {1960},
    publisher = {McGraw-Hill Book Company, Inc.},
    address = {Mineola, New York},
    note = {English translation of the original Russian edition},
    chapter = {5},
    pages = {58--83}
}

@article{Kalensky,
author = {Matthew Kalensky and Stanislav Gordeyev and Matthew R. Kemnetz and Mark F. Spencer},
journal = {J. Opt. Soc. Am. A},
number = {11},
pages = {2163--2174},
publisher = {Optica Publishing Group},
title = {Aero-optical effects, part I. System-level considerations: tutorial},
volume = {41},
month = {Nov},
year = {2024},
url = {https://opg.optica.org/josaa/abstract.cfm?URI=josaa-41-11-2163},
note = {[doi:10.1364/JOSAA.533763]},
}

@article{Gerard,
  title={Fast coherent differential imaging on ground-based telescopes using the self-coherent camera},
  author={Gerard, Benjamin L and Marois, Christian and Galicher, Rapha{\"e}l},
  journal={The Astronomical Journal},
  volume={156},
  number={3},
  pages={106},
  year={2018},
  month = {aug},
  publisher={IOP Publishing},
note = {[doi:10.3847/1538-3881/aad23e]}
}

@inbook{Oppenheim,
author = {Oppenheim, Alan V. and Schafer, Ronald W. and Buck, John R.},
address = {Upper Saddle River, NJ},
booktitle = {Discrete-time signal processing},
chapter = {9--10},
pages = {629-774},
edition = {2nd},
isbn = {0137549202},
language = {eng},
lccn = {98050398},
publisher = {Prentice Hall},
series = {Prentice Hall signal processing series},
title = {Discrete-time signal processing},
year = {1999},
}

@article{PoyneerExperimental,
author = {Poyneer, Lisa and van Dam, Marcos and Véran, Jean-Pierre},
address = {WASHINGTON},
copyright = {Copyright 2015 Elsevier B.V., All rights reserved.},
issn = {1084-7529},
journal = {J. Opt. Soc. Am. A},
language = {eng},
number = {4},
pages = {833--846},
publisher = {Optica Publishing Group},
title = {Experimental verification of the frozen flow atmospheric turbulence assumption with use of astronomical adaptive optics telemetry},
volume = {26},
month = {Apr},
year = {2009},
note = {[doi:10.1364/JOSAA.26.000833]}
}

@article{PoyneerLaboratory,
author = {Poyneer, Lisa A. and Ammons, S. Mark and Kim, Mike K. and Bauman, Brian and Terrel-Perez, Jesse and Lemmer, Aaron J. and Nguyen, Jayke},
address = {WASHINGTON},
copyright = {Copyright 2023 Elsevier B.V., All rights reserved.},
issn = {1559-128X},
journal = {Appl. Opt.},
language = {eng},
number = {8},
pages = {1871--1885},
publisher = {Optica Publishing Group},
title = {Laboratory demonstration of the prediction of wind-blown turbulence by adaptive optics at 8 kHz with use of LQG control},
volume = {62},
month = {Mar},
year = {2023},
note = {[doi:10.1364/AO.474730]}
}

@article{Ziad,
author = {Ziad, Aziz and Schöck, Matthias and Chanan, Gary A. and Troy, Mitchell and Dekany, Richard and Lane, Benjamin F. and Borgnino, Julien and Martin, François},
address = {WASHINGTON},
copyright = {Copyright 2018 Elsevier B.V., All rights reserved.},
issn = {1559-128X},
journal = {Appl. Opt.},
language = {eng},
number = {11},
pages = {2316--2324},
publisher = {Optica Publishing Group},
title = {Comparison of measurements of the outer scale of turbulence by three different techniques},
volume = {43},
month = {Apr},
year = {2004},
note = {[doi:10.1364/AO.43.002316]}
}

@article{ZiadFrom,
author = {Ziad, Aziz and Conan, Rodolphe and Tokovinin, Andrei and Martin, François and Borgnino, Julien},
address = {WASHINGTON},
copyright = {Copyright 2017 Elsevier B.V., All rights reserved.},
issn = {1559-128X},
journal = {Appl. Opt.},
language = {eng},
number = {30},
pages = {5415--5425},
publisher = {Optica Publishing Group},
title = {From the grating scale monitor to the generalized seeing monitor},
volume = {39},
month = {Oct},
year = {2000},
note = {[doi:10.1364/AO.39.005415]}
}

@article{Schöck,
author = {Schöck, Matthias and Mignant, David Le and Chanan, Gary A. and Wizinowich, Peter L. and Dam, Marcos A.Van},
address = {WASHINGTON},
copyright = {Copyright 2017 Elsevier B.V., All rights reserved.},
issn = {1559-128X},
journal = {Appl. Opt.},
language = {eng},
number = {19},
pages = {3705--3720},
publisher = {Optica Publishing Group},
title = {Atmospheric turbulence characterization with the keck adaptive optics systems. I. Open-loop data},
volume = {42},
month = {Jul},
year = {2003},
note = {[doi:10.1364/AO.42.003705]}
}

@article{Avila,
author = {Avila, Remy and Ziad, Aziz and Borgnino, Julien and Martin, François and Agabi, Abdelkrim and Tokovinin, Andrey},
address = {WASHINGTON},
copyright = {Copyright 2017 Elsevier B.V., All rights reserved.},
issn = {1084-7529},
journal = {J. Opt. Soc. Am. A},
language = {eng},
number = {11},
pages = {3070-3082},
publisher = {Optica Publishing Group},
title = {Theoretical spatiotemporal analysis of angle of arrival induced by atmospheric turbulence as observed with the grating scale monitor experiment},
volume = {14},
month = {Nov},
year = {1997},
note = {[doi:10.1364/JOSAA.14.003070]}
}

@article{Fried,
number = {10},
pages = {1372--1379},
publisher = {Optica Publishing Group},
title = {Optical Resolution Through a Randomly Inhomogeneous Medium for Very Long and Very Short Exposures},
volume = {56},
month = {Oct},
year = {1966},
author = {Fried, D. L.},
address = {WOODBURY},
issn = {0030-3941},
journal = {J. Opt. Soc. Am.},
language = {eng ; jpn},
note = {[doi:10.1364/JOSA.56.001372]}
}

@article{Coulman,
number = {1},
pages = {155--160},
publisher = {Optica Publishing Group},
title = {Outer scale of turbulence appropriate to modeling refractive-index structure profiles},
volume = {27},
month = {Jan},
year = {1988},
author = {C. E. Coulman and J. Vernin and Y. Coqueugniot and J. L. Caccia},
address = {WASHINGTON},
issn = {0003-6935},
journal = {Appl. Opt.},
language = {eng},
note = {[doi:10.1364/AO.27.000155]}
}

@article{Lin,
  title={Design considerations of photonic lanterns for diffraction-limited spectrometry},
  author={Lin, Jonathan and Jovanovic, Nemanja and Fitzgerald, Michael P},
  journal={J. Opt. Soc. Am. B},
  volume={38},
  number={7},
  pages={A51--A63},
  month = {Jul},
  year={2021},
  publisher={Optica Publishing Group},
note = {[doi:10.1364/JOSAB.423664]}
}

@article{Dayton,
  title={Scaled-laboratory demonstrations of deep-turbulence conditions},
  author={Dayton, David C and Spencer, Mark F},
  journal={Appl. Opt.},
  volume={63},
  number={16},
  pages={E54--E63},
  year={2024},
  month = {Jun},
  publisher={Optica Publishing Group},
note = {[doi:10.1364/AO.520208]}
}

@article{Jia,
  title={Digital twin of atmospheric turbulence phase screens based on deep neural networks},
  author={Jia, Peng and Wang, Weihua and Ning, Runyu and Xue, Xiaolei},
  journal = {Opt. Express},
  volume={30},
  number={12},
  pages={21362--21376},
  year={2022},
  month = {Jun},
  publisher={Optica Publishing Group},
note = {[doi:10.1364/OE.460244]}
}

@article{Boddeda,
  title={Achievable capacity of geostationary-ground optical links},
  author={Boddeda, Rajiv and Arrieta, Daniel Romero and Almonacil, Sylvain and Renaudier, J{\'e}r{\'e}mie and Bigo, S{\'e}bastien},
  journal={Journal of Lightwave Technology},
  volume={41},
  number={12},
  pages={3717--3725},
  year={2023},
  publisher={IEEE},
note = {[doi:10.1109/JLT.2023.3271824]}
}

@inproceedings{Sheikh,
  title={Dynamic DH-MBIR for phase-error estimation from streaming digital-holography data},
  author={Sheikh, Ali G and Pellizzari, Casey J and Kisner, Sherman J and Buzzard, Gregery T and Bouman, Charles A},
  booktitle={2023 57th Asilomar Conference on Signals, Systems, and Computers},
  pages={784-788},
  year={2023},
  organization={IEEE},
  note = {[doi:10.1109/IEEECONF59524.2023.10477047]}
}

@inproceedings{Snyder,
author = {Adam Snyder and Srikar Srinath and Bruce Macintosh and Aaron Roodman},
title = {{Temporal characterization of Zernike decomposition of atmospheric turbulence}},
volume = {9906},
booktitle = {Ground-based and Airborne Telescopes VI},
editor = {Helen J. Hall and Roberto Gilmozzi and Heather K. Marshall},
series = {Proc. SPIE},
pages = {990642},
year = {2016},
note = {[doi:10.1117/12.2234362]},
URL = {https://doi.org/10.1117/12.2234362}
}

@inproceedings{Sridhar,
author = {Venkatesh Sridhar and Sherman J. Kisner and Samuel P. Midkiff and Charles A. Bouman},
title = {{Fast algorithms for model-based imaging through turbulence}},
volume = {11543},
booktitle = {Artificial Intelligence and Machine Learning in Defense Applications II},
editor = {Judith Dijk},
series = {Proc. SPIE},
pages = {1154304},
year = {2020},
note = {[doi:10.1117/12.2570789]},
URL = {https://doi.org/10.1117/12.2570789}
}

@inproceedings{Lehtonen,
author = {Jonatan Lehtonen and Carlos M. Correia and Tapio Helin},
title = {{Limits of turbulence and outer scale profiling with non-Kolmogorov statistics}},
volume = {10703},
booktitle = {Adaptive Optics Systems VI},
editor = {Laird M. Close and Laura Schreiber and Dirk Schmidt},
series = {Proc. SPIE},
pages = {107036C},
year = {2018},
note = {[doi:10.1117/12.2313960]},
URL = {https://doi.org/10.1117/12.2313960}
}

@inproceedings{eBraga,
  title={Coherence length measurements under strong scintillation conditions using a five-layer laboratory-scaled atmospheric simulator},
  author={Alexandre de Pinho e Braga and Denis W. Oesch and David C. Dayton and Mark F. Spencer},
  booktitle={Conference on Lasers and Electro-Optics},
  pages={SW4E.7},
  year={2020},
  publisher={Optica Publishing Group},
  journal = {Conference on Lasers and Electro-Optics},
    note = {[doi:10.1364/CLEO\_SI.2020.SW4E.7]}
}

@inproceedings{BurnsEstimation,
author = {Burns, Robert and Gordeyev, Stanislav and Jumper, Eric and Gogineni, Sivaram and Paul, Michael and Wittich, Donald J.},
booktitle = {52nd AIAA Aerospace Sciences Meeting},
copyright = {Copyright 2014 Elsevier B.V., All rights reserved.},
isbn = {9781624102561},
language = {eng},
title = {Estimation of Aero-Optical Wavefronts Using Optical and Non-Optical Measurements},
pages={0319},
year = {2014},
publisher = {American Institute of Aeronautics and Astronautics},
note = {[doi:10.2514/6.2014-0319]}
}

@article{JumperAAOL,
author = {Eric J. Jumper and Michael A. Zenk and Stanislav V. Gordeyev and David A. Cavalieri and Matthew Whitely},
title = {{Airborne Aero-Optics Laboratory}},
volume = {52},
journal = {Optical Engineering},
number = {7},
publisher = {SPIE},
pages = {071408},
year = {2013},
note = {[doi:10.1117/1.OE.52.7.071408]},
URL = {https://doi.org/10.1117/1.OE.52.7.071408}
}

@inbook{Geary,
author = {Geary, Joseph M},
volume={TT18},
address = {Bellingham, Washington},
booktitle = {Introduction to wavefront sensors},
isbn = {1-61583-744-2},
language = {eng},
publisher = {SPIE Optical Engineering Press},
series = {Tutorial texts in optical engineering},
title = {Introduction to wavefront sensors },
year = {1995},
chapter = {5-7},
pages = {53--103}
}

@article{Rasouli,
  title={Investigation of the anisotropy and scaling of the phase structure function of a spatially coherent light beam propagating through convective air turbulence},
  author={Rasouli, Saifollah and Mohammadi Razi, Ebrahim and Niemela, JJ},
  journal = {J. Opt. Soc. Am. A},
  volume={39},
  number={9},
  pages={1641--1649},
  month = {Sep},
  year={2022},
  publisher={Optica Publishing Group},
note = {[doi:10.1364/JOSAA.464285]}
}

@article{Silbaugh,
  title={Characterization of atmospheric turbulence phase statistics using wave-front slope measurements},
  author={Silbaugh, Eric E and Welsh, Byron M and Roggemann, Michael C},
  journal = {J. Opt. Soc. Am. A},
  volume={13},
  number={12},
  pages={2453--2460},
  month = {Dec},
  year={1996},
  publisher = {Optica Publishing Group},
note = {[doi:10.1364/JOSAA.13.002453]}
}

@inproceedings{UtleyBoiling,
author = {Jeffrey W. Utley and Gregery T. Buzzard and Charles A. Bouman and Matthew R. Kemnetz},
title = {{Boiling flow parameter estimation from boundary layer data}},
volume = {13619},
booktitle = {Unconventional Imaging, Sensing, and Adaptive Optics 2025},
editor = {Jean J. Dolne and Santasri R. Bose-Pillai and Matthew Kalensky},
series = {Proc. SPIE},
pages = {136190L},
year = {2025},
note = {[doi:10.1117/12.3063655]},
URL = {https://doi.org/10.1117/12.3063655}
}

@article{KemnetzAnalysis,
author = {Kemnetz, Matthew R. and Gordeyev, Stanislav},
title = {Analysis of Aero-Optical Jitter in Convective Turbulent Flows Using Stitching Method},
journal = {AIAA Journal},
volume = {60},
number = {1},
pages = {14-30},
year = {2022},
note = {[doi:10.2514/1.J060756]},

URL = { 
    
        https://doi.org/10.2514/1.J060756
    
    

},
eprint = { 
    
        https://doi.org/10.2514/1.J060756
    
    

}
}

@inbook{Tyson,
author = {Tyson, Robert K},
title = {Introduction to Adaptive Optics},
address = {Bellingham, Washington},
copyright = {2000},
isbn = {0819435112},
language = {eng},
chapter = {4},
pages = {33--44},
organization = {International Society for Optics and Photonics},
publisher = {SPIE},
year = {2000},
note = {[doi:10.1117/3.358220]}
}

@article{Faghihi,
author = {Azin Faghihi and Jonathan Tesch and Steve Gibson},
title = {{Identified state-space prediction model for aero-optical wavefronts}},
volume = {52},
journal = {Optical Engineering},
number = {7},
publisher = {SPIE},
pages = {071419},
year = {2013},
note = {[doi:10.1117/1.OE.52.7.071419]},
URL = {https://doi.org/10.1117/1.OE.52.7.071419}
}

@article{Taylor,
  title={The spectrum of turbulence},
  author={Taylor, Geoffrey Ingram},
  journal={Proceedings of the Royal Society of London. Series A-Mathematical and Physical Sciences},
  volume={164},
  number={919},
  pages={476-490},
  year={1938},
  publisher={The Royal Society},
note = {[doi:10.1098/rspa.1938.0032]}
}

@inproceedings{Lloyd,
author = {Robert L. Lloyd and Tyler J. Hardy and Mark F. Spencer and Casey J. Pellizzari},
title = {{Dynamic image correction and wavefront sensing with a digital holographic sensor using 4D implicit neural representations}},
volume = {13619},
booktitle = {Unconventional Imaging, Sensing, and Adaptive Optics 2025},
editor = {Jean J. Dolne and Santasri R. Bose-Pillai and Matthew Kalensky},
series = {Proc. SPIE},
pages = {136190Y},
year = {2025},
note = {[doi:10.1117/12.3063178]},
URL = {https://doi.org/10.1117/12.3063178}
}

@inproceedings{SheikhDynamic,
author = {Ali G. Sheikh and Casey J. Pellizzari and Sherman J. Kisner and Gregery T. Buzzard and Charles A.  Bouman},
title = {{Dynamic DH-MBIR for low-latency wavefront estimation in the presence of atmospheric boiling}},
volume = {12693},
booktitle = {Unconventional Imaging, Sensing, and Adaptive Optics 2023},
editor = {Jean J. Dolne and Mark F. Spencer and Santasri R. Bose-Pillai},
series = {Proc. SPIE},
pages = {1269308},
year = {2023},
note = {[doi:10.1117/12.2676406]},
URL = {https://doi.org/10.1117/12.2676406}
}

@phdthesis{KemnetzDissertation,
  author  = "Matthew Kemnetz",
  title   = "Analysis of the Aero-Optical Component of the Jitter Using the Stitching Method",
  school  = "University of Notre Dame",
  year    = "2019",
  month   = "July",
note = {[doi:10.7274/k930bv76j74]}
}

@article{Andrade,
  title={Estimation of atmospheric turbulence parameters from Shack--Hartmann wavefront sensor measurements},
  author={Andrade, Paulo P and Garcia, Paulo JV and Correia, Carlos M and Kolb, Johann and Carvalho, Maria In{\^e}s},
  journal={Monthly Notices of the Royal Astronomical Society},
  volume={483},
  number={1},
  pages={1192-1201},
  year={2018},
  publisher={Oxford University Press},
  month = {11},
note = {[doi:10.1093/mnras/sty3181]}
}

@misc{Repo,
  author       = {Utley, Jeffrey W and Buzzard, Gregery T and Bouman, Charles A and Kemnetz, Matthew R},
  title        = {Boiling Flow},
  year         = {2025},
  howpublished = {Software library available from \url{https://github.com/jeffreyutley/boiling_flow}},
}
\bibliographystyle{spiejour}

\vspace{2ex}\noindent\textbf{Jeffrey Utley} is a PhD candidate in mathematics at Purdue University. He received his BS degree in mathematics from the University of Tennessee in 2022. His current research focuses on generating synthetic aero-optic phase screens using statistical models.

\vspace{1ex}
\noindent Biographies and photographs of the other authors are not available.

\listoffigures
\listoftables

\end{spacing}
\end{document}